\documentclass[a4paper, 11pt]{article}

\usepackage{fullpage} 
\usepackage{hyperref}
\usepackage{amsmath,amssymb,empheq}
\usepackage{graphicx}
\usepackage[colorinlistoftodos]{todonotes}
\usepackage{float}
\usepackage{wasysym}
\usepackage{lineno}
\usepackage{tcolorbox}
\usepackage{soul}
\usepackage{authblk}

\title{Analytical formulas of coherent-synchrotron-radiation induced microbunching gain and emittance growth in an arbitrary achromatic four-bend chicane}

\author[1]{Bingxi Liu}
\author[1]{Cheng-Ying Tsai}
\author[2]{Yi Jiao}
\author[2]{Weihang Liu}
\author[2]{Fancong Zeng}
\author[3]{Weilun Qin}
\affil[1]{Huazhong University of Science and Technology, Wuhan, 430074, China}
\affil[2]{Institute of High Energy Physics, Chinese Academy of Sciences, and University of Chinese Academy of Sciences, Beijing 100049, China}
\affil[3]{Deutsches Elektronen-Synchrotron DESY, Notkestr. 85, 22607 Hamburg, Germany}

\begin{document}

\maketitle

The coherent synchrotron radiations (CSR) emitted by a high-brightness electron beam during transport in a bending magnet is a double-edged sword in electron accelerators. While CSR contributes to a stronger radiation field than the incoherent radiation, it simultaneously leads to degradation of the electron beam quality. Specifically, CSR effects manifest in increases of the beam energy spread and the projected emittance, and amplification of the microbunching instability. A dedicated design of the multi-bend transport lines to mitigate CSR effects has recently become a crucial consideration in modern high-brightness electron accelerators. This paper presents analytical formulas for the CSR-induced microbunching instability gain and for the induced emittance growth in an arbitrary achromatic four-bend chicane with inclusion of both the steady-state and transient CSR effects. For the microbunching instability, an iterative method is employed to solve the integral equation for the bunching factor, providing satisfactory gain formulas. Regarding the CSR-induced emittance growth, based on the linear transfer matrices and incorporating the expressions of the CSR-induced energy spread, analytical formulas are derived for the projected emittance growth in an arbitrary four-bend transport section. The analytical formulas are compared and show good agreement with semi-analytical Vlasov calculations and particle tracking simulations. As an application, the obtained analytical formulas are applied to evaluate the CSR effects in the design of a general achromatic four-bend bunch compressor chicane, providing a quick estimate on the microbunching gain and the induced emittance growth. From the widely adopted symmetric C-shape chicane to a non-symmetric S-shape chicane, a design recently proposed by the co-authors, our analytical formulas offer insight into the evolution of the microbunching gain and the emittance growth with the variations of design parameters. In comparison to the time-consuming, full numerical particle tracking simulations currently employed for CSR effect analyses, the analytical formulas presented in this paper significantly reduce the evaluation time, enabling systematic study of parametric dependencies and comprehensive optimization with inclusion of CSR effects within specified design parameter ranges.






\section{Introduction}\label{SecI}

The brightness of a charged particle beam can be characterized by the volume it occupies in phase space. Generally, a high-brightness electron beam exhibits low beam emittance, small energy spread, and high bunch current. The synchrotron radiation generated during the bending process of such a high-brightness electron bunch, when the wavelength component is comparable to the bunch length or the scale of density modulations within the bunch, is often referred to as the coherent synchrotron radiation (CSR). CSR can induce correlations between different longitudinal slices within the bunch, leading to collective effects. The CSR-induced energy deviation differs in the longitudinal slices of the bunch in and after the bending magnet. In downstream regions with non-zero dispersion during transport, CSR will result in the projected beam emittance growth~\cite{Ref01,Ref02,Ref03,Ref04,Ref05} and microbunching instability (MBI)~\cite{Ref06,Ref07,Ref08,Ref09,Ref10}. To the authors' knowledge, most existing literature simplifies the study of CSR effects by assuming only presence of steady-state CSR fields in theoretical models for convenience. Evaluation of the transient CSR effects is generally performed through numerical methods, for example particle tracking simulations. In sum, effectively suppressing or controlling CSR effects has become one of the most challenging issues in the design of high-brightness electron beam transport systems in recent years.

From the perspective of the kinetic theory of collective effects in high-brightness electron beams, typically developed based on the Vlasov equation, the analysis of microbunching instability generally adopts a semi-analytical approach. This approach transforms the linearized Vlasov equation into an integral equation and solves for the bunching factor (which we shall define soon below), allowing for the proper inclusion of the transverse effects of beam intrinsic spreads. The theory of microbunching in a single-pass system was early developed by Saldin et al.~\cite{Ref07}, Heifets et al.~\cite{Ref08}, and Huang and Kim~\cite{Ref09}. The theoretical model has been continuously refined. Saldin et al. treated this problem as a klystron-like instability, considered the case without bunch compression, and assumed the high-gain approximation. This theory revealed the fundamental physical picture of microbunching instability. Subsequently, Heifets et al. extended the treatment to include bunch compression and a finite transverse beam emittance. Applying the standard perturbation technique to linearize the Vlasov equation and integrating along the unperturbed characteristics, Heifets et al. derived a linear integral equation in terms of the bunching factor. The authors in Ref.~\cite{Ref08} solved the integral equation numerically, a method applicable to a general linear lattice. For a typical, symmetric three-bend magnetic bunch compressor chicane commonly used in linear accelerators, the authors in Ref.~\cite{Ref09} derived an analytical formula using an iterative approach to estimate the steady-state CSR microbunching gain. The above work was largely based on the linearized Vlasov equation and assumed constant energy along the beam transport line. The above methods are semi-analytical or analytical, and include only steady-state CSR effects. The main approaches for numerical analysis of collective effects induced by CSR are the intuitive but computationally expensive and resource-intensive particle tracking simulation, and the direct solution of the low-dimensional Vlasov equation. Particle tracking simulation is subject to numerical noise due to the very limited number of simulation particles. A direct solution of the (nonlinear) Vlasov equation for the longitudinal phase-space distribution only, based on the semi-Lagrangian approach, was studied by Venturini et al.~\cite{Ref10} This approach resolves the issue of numerical noise, but the transverse effects of beam intrinsic spread are only approximately accounted for. Still, the above considerations of CSR effects assume steady-state effects. Our recently developed six-dimensional semi-analytical Vlasov solver~\cite{Ref11,Ref12} further extends the capabilities by including richer high-frequency impedance models, such as the non-ultrarelativistic CSR impedances in the relatively low-energy regime~\cite{Ref13} and the exit transient CSR or drift CSR impedance at the end of bending magnets~\cite{Ref14}. Recent studies have shown that the impact of the exit transient CSR field can be significant and non-negligible. Additionally, our Vlasov solver allows for transversely coupled beams or coupled lattices and includes intrabeam scattering effects~\cite{Ref15,Ref16}. It is well known that particle tracking simulation is the mainstream method for beam dynamics studies, but suffers from the drawbacks of being time-consuming and requiring massive computational resources. Being approximately three orders of magnitude faster, this semi-analytical tool enables efficient estimation of the MBI for a given linear lattice, helping lattice design by quickly assessing the lattice performance.

Before delving into the two concepts that will be discussed in this paper, the CSR-induced microbunching instability (MBI) and the emittance growth, it would be beneficial to briefly review some of the recent efforts within the accelerator community to mitigate MBI and suppress emittance growth. The theory of MBI has clearly demonstrated that the involvement of transverse beam emittances and/or energy spread, leading to Landau damping, is an effective stabilizing mechanism for MBI. Laser heating has been widely employed in linac-based free-electron laser (FEL) facilities to effectively suppress MBI~\cite{Ref17,Ref18}. The underlying principle is to utilize a short-wavelength laser interacting with electrons in a short undulator to induce an additional uncorrelated energy spread, thereby enhancing Landau damping. In addition to the laser heating technique, the electron-magnetic-phase mixing approach has also been proposed, where a magnetically mixing chicane is employed to smooth the bunch current and energy distribution by forcing the electrons to slip in the longitudinal phase space~\cite{Ref19}. Experimental studies have demonstrated that this technique can reduce MBI by an order of magnitude~\cite{Ref19}. Other specialized beamline designs have also been proposed to mitigate MBI. For instance, introducing a set of transverse deflecting RF cavities upstream and downstream of a bunch compressor can increase additional Landau damping and thus effectively suppress MBI~\cite{Ref20}. The first RF cavity introduces an energy spread to suppress microbunching instability, while the second RF cavity eliminates the induced energy spread. Moreover, in Ref.~\cite{Ref21}, Qiang et al. proposed a cost-effective scheme to suppress MBI in a linac-based X-ray free-electron laser light source by inserting a pair of bending magnets in the accelerator transport system.

In contrast to increasing the bunch energy spread to induce stronger Landau damping, a less intuitive approach is to employ optical balance, where the local momentum compaction function of the transport line is minimized as much as possible, thereby satisfying local isochronicity, which can effectively suppress CSR-induced MBI. However, different transport lattices have varying optics functions, and for multi-bend transport lines, conducting a comprehensive and systematic study using particle tracking simulations would be extremely time-consuming and almost unlikely to implement. Leveraging the advantages of our developed semi-analytical tool, we can now systematically explore various schemes for mitigating MBI and unravel the underlying physical mechanisms of microbunching amplification. For example, thanks to the tool, Refs.~\cite{Ref22,Ref2202} provide generic conditions for suppressing MBI in multi-bend transport lines. It is worth mentioning that in recent experimental work at FERMI, they explored and validated an optical-based approach, where manipulating a few upstream quadrupoles to modify the downstream transport line's dispersion invariant function allowed for controlling the Landau damping strength, thereby achieving MBI suppression~\cite{Ref23}.

Although the semi-analytical calculation offers a speedup of nearly three orders of magnitude compared to particle tracking simulations, enabling efficient estimations, the semi-analytical Vlasov solver still requires some time for initial rough optimization during the design phase. As a primary contribution of this paper, we will derive some analytical formulas specifically for the achromatic four-bend chicane transport lattice, which can be directly used to estimate the MBI induced by CSR (including both steady-state and transient effects). The analytical formulas themselves maximize the efficiency of numerical computations, enabling a comprehensive exploration of multi-parameter dependencies and making it feasible to perform a comprehensive multi-parameter optimization.

In addition to the CSR-induced MBI in the longitudinal dimension, CSR can also potentially lead to emittance growth in the transverse dimension. Based on early insights into the one-dimensional steady-state CSR field, Douglas~\cite{Ref24,Ref2402,Ref2403,Ref2404} proposed the concept of cell-to-cell phase matching to compensate or cancel the CSR kicks. Subsequently, this concept was further adopted and explored, such as by S. Di Mitri et al., who employed cell-to-cell phase matching to compensate or cancel the CSR kicks~\cite{Ref25}. Other approaches include Hajima et al., who used the beam envelope matching method by characterizing the transverse phase-space ellipse tilt due to CSR~\cite{Ref26}. This method concludes that with proper arrangement of lattice optics in a unit cell along the major axis of the transverse beam phase space ellipse, the beam emittance growth due to steady-state CSR can be minimized. With dedicated beamline design, this approach can achieve cancellation of CSR-induced emittance growth. Following the phase matching and beam envelope methods, Jiao et al. proposed a CSR point-kick model and extended the above two methods to provide generic conditions for suppression of the CSR-induced emittance growth in a double-bend achromat unit~\cite{Ref27}. These schemes have been primarily applied to isochronous transport lattices, where the bunch length remains approximately constant during the transport process. In recent years, these suppression schemes have begun to be extended to cases allowing bunch length variation, such as application to bunch compressor lattices. For example, Di Mitri and Cornacchia extended their cell-to-cell phase matching analysis to the case of compressor arcs~\cite{Ref28}. Following Jiao et al., further studies based on the optical balance method have been successfully applied to many compressor systems, e.g., double-bend achromats (DBA) and triple-bend achromat (TBA) compressors~\cite{Ref29,Ref30,Ref31}. Again, these theoretical models studying the suppression schemes of CSR-induced emittance growth simply assume only steady-state CSR. When the impact of the exit transient CSR field becomes significant, it is no longer negligible.

In addition to deriving an analytical formula for the CSR-induced MBI, in this paper we will also specifically derive a formula for the beam projected emittance growth induced by CSR for the achromatic four-bend chicane, using the linear optics method. Here we emphasize that the existing formulas only accounted for the steady-state CSR effects, while the analytical formula derived in this paper includes both the steady-state and the transient CSR effects.

Here, we briefly summarize the main contributions of this paper. For the achromatic four-bend chicane, we have derived analytical formulas that can be directly used to estimate the MBI gain and the projected emittance growth induced by both steady-state and transient CSR effects. Taking advantage of these analytical formulas, we attempt to perform a quick scan of relevant parameters and evaluate the MBI gain and emittance growth provided with two example lattices. We expect that the analytical formulas can provide insights into the lattice design aimed at controlling CSR-induced microbunching as well as emittance growth. These analytical formulas are inherently well-suited for preliminary design optimization of a four-dipole chicane transport lattice.

This paper is organized as follows. In Sec.~\ref{SecII}, we first lay the groundwork for deriving the MBI gain formula and the emittance growth formula by preparing the necessary ingredients, including the transfer matrix elements based on the linear optics for the four-bend chicane (Sec.~\ref{SecIIA}), as well as the CSR-related characteristics, particularly the expressions for the CSR-induced energy spread increase (Sec.~\ref{SecIIB}). Armed with these ingredients, we will then proceed to derive the formula for the CSR-induced MBI gain in Sec.~\ref{SecIIC}, followed by the derivation of the formula for CSR-induced emittance growth in Sec.~\ref{SecIID}. For brevity, we have relegated some details of mathematical derivation to the Appendix. As an application of the derived formulas, in Sec.~\ref{SecIII}, we present two examples, a symmetric C-shape chicane and a non-symmetric S-shape chicane, where we employ the derived analytical formulas to estimate their CSR-induced effects. Finally, we summarize the work presented in this paper in Sec.~\ref{SecIV}.

\section{Theoretical formulation}\label{SecII}

\begin{figure}[htbp]
\centering
\includegraphics[width=14cm]{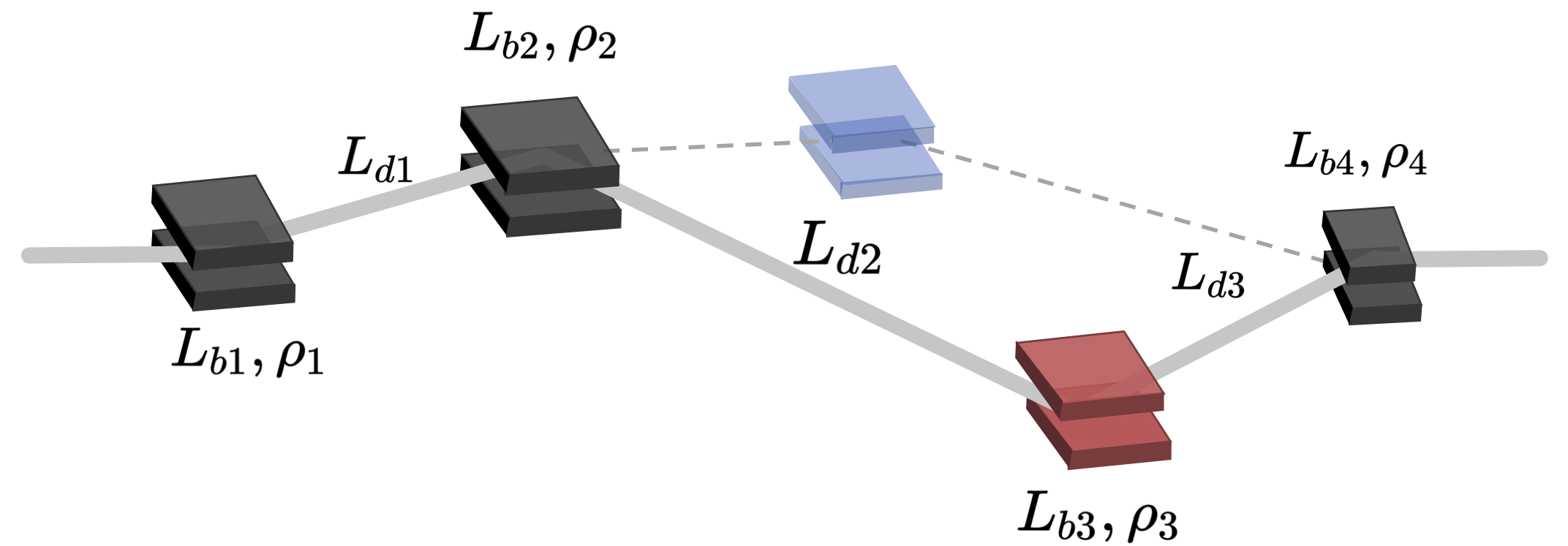}
\caption{\textmd{Schematic layout of four-bend transport line. We assumed that the particle beam travels from left to right. When the third bending magnet is placed on the same side as the second one, we call it the C-shape chicane (blue); when placed on the opposite side, we dub it as the S-shape chicane (red).}}
\label{Fig1}
\end{figure}

Before embarking on the derivation of the formulas, let us define the four-bend chicane and introduce some notations that will be used. The four-bend chicane, as illustrated in Fig.~\ref{Fig1}, consists of four bending magnets with lengths denoted as $L_{b1}, L_{b2}, L_{b3}, L_{b4}$, and the corresponding bending radii denoted as $\rho_{1}, \rho_{2}, \rho_{3}, \rho_{4}$. For the geometric configuration shown in Fig.~\ref{Fig1}, referred to as an S-shape chicane (with the third bend marked red), we adopt the convention $\rho_{i} > 0~(i=1,2,3,4)$. For a C-shape chicane, where the third bend is marked as red, we have $(\rho_{1,2} > 0, \rho_{3,4} < 0)$. The lengths of the three drift sections between the four bending magnets are denoted as $L_{d1}, L_{d2}, L_{d3}$, respectively. Within each bending magnet, the individual path-length coordinates are written as $s_1, s_2, s_3, s_4$, with integration ranging from $0$ to $L_{b1}, L_{b2}, L_{b3}, L_{b4}$, respectively. In the three drift sections, the path-length coordinates are denoted as $\zeta_1, \zeta_2, \zeta_3$, with integration ranging from 0 to $L_{d1}, L_{d2}, L_{d3}$, respectively. The total length of the lattice is $L_T = \sum_{j=1}^{4} L_{bj} + \sum_{j=1}^{3} L_{dj}$, and the path-length coordinates at the entrance and exit of the lattice are denoted as $s_0 = 0$ and $s_f = L_T$.

In our subsequent derivations, we will assume that this four-bend chicane is achromatic, implying that the overall transport section satisfies $R_{16}(s_f) = 0, R_{26}(s_f) = 0$, which is in accord with the coaxial condition depicted in Fig.~\ref{Fig1} for the given geometric configuration.

\subsection{Pure optics: transfer matrix elements}\label{SecIIA}

The transfer matrix is the most common method used to describe pure linear optics in absence of collective interactions among beam particles. In what follows, we discuss the beam dynamics in the 4-D phase space coordinates, written as $(x, x', z, \delta)^T$, where $x$ and $x'$ are, respectively, the transverse offset and divergence of a particle within the beam with respect to the reference particle, $z$ is the longitudinal deviation of a particle within the beam from the central particle ($z > 0$ indicates the bunch head), and $\delta$ is the relative energy deviation of a particle within the beam from the reference beam energy. Relative to the starting point $s = s_0 = 0$, the $4 \times 4$ transfer matrix at the location $s$ can be written as $\mathbf{R}(s) = \mathbf{R}(0 \to s)$, which can be constructed by the sequential multiplication of the transfer matrices for each component. The transfer matrix of from a certain location $\tau$ to $s$ in the transport can be obtained through $\mathbf{R}(\tau \to s) = \mathbf{R}(0 \to s) \mathbf{R}^{-1}(0 \to \tau)$. The following derivation assumes that the length of the bending magnets is much shorter than the drift length, i.e., $L_{b} \ll L_{d}$, so that the transfer matrix of the bending magnets can be approximated by the thin-lens approximation. Here, we follow the convention in the accelerator community and name each element of the $4 \times 4$ transfer matrix in the following form
\begin{equation}\label{Eq1}
\left(\begin{array}{cccc}
R_{11} & R_{12} & 0 & R_{16} \\
R_{21} & R_{22} & 0 & R_{26} \\
R_{51} & R_{52} & 1 & R_{56} \\
0 & 0 & 0 & 1
\end{array}\right),
\end{equation}
where the elements $R_{15}, R_{25}, R_{61}, R_{62}, R_{65}$ are considered vanishing elements.

In this section, we will summarize the transfer matrix elements based on pure optics that will be used later. First, we need the transfer matrices within the four bending magnets. For the first bending magnet,
\begin{equation}\label{Eq2}
\mathbf{R}\left(s_1\right)=\mathbf{R}\left(0 \to s_1\right)=\left(\begin{array}{cccc}
1 & s_1 & 0 & \frac{s_1^2}{2 \rho_1} \\
0 & 1 & 0 & \frac{s_1}{\rho_1} \\
-\frac{s_1}{\rho_1} & -\frac{s_1^2}{2 \rho_1} & 1 & -\frac{s_1^3}{6 \rho_1^2} \\
0 & 0 & 0 & 1
\end{array}\right),
\end{equation}
here $0 \le s_1 \le L_{b1}$ and $\rho_1$ being the bending radius. For the second bending magnet,
\begin{equation}\label{Eq3}
\mathbf{R}\left(s_2\right)=\mathbf{R}\left(0 \to s_2\right)=\left(\begin{array}{cccc}
1 & L_{b 1}+L_{d 1}+s_2 & 0 & \frac{L_{b 1} L_{d 1}}{\rho_1} \\
0 & 1 & 0 & \frac{L_{b 1}}{\rho_1}-\frac{s_2}{\rho_2} \\
\frac{s_2}{\rho_2}-\frac{L_{b 1}}{\rho_1} & \frac{L_{d 1} s_2}{\rho_2} & 1 & \frac{L_{b 1} L_{d 1} s_2}{\rho_1 \rho_2} \\
0 & 0 & 0 & 1
\end{array}\right),
\end{equation}
here $0 \le s_2 \le L_{b2}$ and $\rho_2$ being the corresponding bending radius of the second one. Since the matrix is evaluated from the very beginning $s_0 = 0$, the matrix elements depend on the parameters of the first magnet and the in-between drift length $L_{d1}$. Similarly, for the third bending magnet,
\scriptsize
\begin{equation}\label{Eq4}
\begin{aligned}
&\mathbf{R}\left(s_3\right)=\mathbf{R}\left(0 \to s_3\right)= \\
&\left(\begin{array}{cccc}
1 & L_{d 1}+L_{d 2}+L_{b 1}+L_{b 2}+s_3 & 0 & \frac{L_{b 1} L_{d 1}}{\rho_1}+\frac{L_{b 1} L_{d 2}}{\rho_1}-\frac{L_{b 2} L_{d 2}}{\rho_2} \\
0 & 1 & 0 & \frac{L_{b 1}}{\rho_1}-\frac{L_{b 2}}{\rho_2}+\frac{s_3}{\rho_3} \\
\frac{L_{b 2}}{\rho_2}-\frac{L_{b 1}}{\rho_1}-\frac{s_3}{\rho_3} & \frac{L_{b 2} L_{d 1}}{\rho_2}-\frac{\left(L_{d 1}+L_{d 2}\right) s_3}{\rho_3} & 1 & \left(\frac{L_{b 2} L_{d 2}}{\rho_2 \rho_3}-\frac{L_{b 1} L_{d 2}}{\rho_1 \rho_3}-\frac{L_{b 1} L_{d 1}}{\rho_1 \rho_3}\right) s_3+\frac{L_{b 1} L_{b 2} L_{d 1}}{\rho_1 \rho_2} \\
0 & 0 & 0 & 1
\end{array}\right),
\end{aligned}
\end{equation}
\normalsize
here $0 \le s_3 \le L_{b3}$, $\rho_3$ being the corresponding bending radius of the third one and $L_{d1}, L_{d2}$ being the drift lengths. For the last bending magnet, we have
\begin{equation}\label{Eq5}
\mathbf{R}\left(s_4\right)=\mathbf{R}\left(0 \to s_4\right)=\left(\begin{array}{cccc}
1 & R_{12}(0 \to s_4) & 0 & R_{16}(0 \to s_4) \\
0 & 1 & 0 & R_{26}(0 \to s_4) \\
R_{51}(0 \to s_4) & R_{52}(0 \to s_4) & 1 & R_{56}(0 \to s_4) \\
0 & 0 & 0 & 1
\end{array}\right), 
\end{equation}
where the shorthand notations represent
\begin{equation}\label{Eq6}
\begin{aligned}
R_{12}(0 \to s_4) &= L_{d 1}+L_{d 2}+L_{d 3}+L_{b 1}+L_{b 2}+L_{b 3}+s_4 \\
R_{16}(0 \to s_4) &= \frac{L_{b 1}\left(L_{d 1}+L_{d 2}+L_{d 3}\right)}{\rho_1}+\frac{L_{b 3} L_{d 3}}{\rho_3}-\frac{L_{b 2}\left(L_{d 2}+L_{d 3}\right)}{\rho_2} \\
R_{26}(0 \to s_4) &= \frac{L_{b 1}}{\rho_1}-\frac{L_{b 2}}{\rho_2}+\frac{L_{b 3}}{\rho_3}-\frac{s_4}{\rho_4} \\
R_{51}(0 \to s_4) &=\frac{L_{b 2}}{\rho_2}-\frac{L_{b 1}}{\rho_1}-\frac{L_{b 3}}{\rho_3}+\frac{s_4}{\rho_4} = \frac{s_{4-} L_{b 4}}{\rho_4}\\
R_{52}(0 \to s_4) &= \frac{\left(L_{d 1}+L_{d 2}+L_{d 3}\right)\left(s_4-L_{b 4}\right)}{\rho_4} \\
R_{56}(0 \to s_4) &= \frac{L_{b 1} L_{b 2} L_{d 1}}{\rho_1 \rho_2}+\frac{L_{b 2} L_{b 3} L_{d 2}}{\rho_2 \rho_3}-\frac{L_{b 1} L_{b 3} L_{d 2}}{\rho_1 \rho_3}-\frac{L_{b 1} L_{b 3} L_{d 1}}{\rho_1 \rho_3} \equiv R_{56}
\end{aligned}
\end{equation}
where the aforementioned achromatic conditions $R_{16}(0 \to s_f) = 0, R_{26}(0 \to s_f) = 0$ have been assured. Moreover, the symplectic condition will require that $R_{51}(0 \to s_f) = 0, R_{52}(0 \to s_f) = 0$. Here we note that, in the above expression, the longitudinal momentum compaction function of the entire four-bend chicane $R_{56}(L_T) = R_{56}(0 \to s_f)$ is simply denoted as $R_{56}$.

In addition to the transfer matrix elements within the bending magnets, when estimating the impact of the exit-transient CSR fields on the beam, we need the transfer matrix elements in the downstream drift sections, which are
\begin{equation}\label{Eq7}
R_{16}\left(\zeta_1\right)=\frac{L_{b 1}}{\rho_1}\zeta_1, \quad R_{16}\left(\zeta_2\right)=\frac{L_{b 1} L_{d 1}}{\rho_1}+\left(\frac{L_{b 1}}{\rho_1}-\frac{L_{b 2}}{\rho_2}\right) \zeta_2,
\end{equation}
\begin{equation}\label{Eq8}
R_{16}\left(\zeta_3\right)=\frac{L_{b 1} L_{d 1}}{\rho_1}+\frac{L_{b 1} L_{d 2}}{\rho_2}-\frac{L_{b 2} L_{d 2}}{\rho_2}+\left(\frac{L_{b 1}}{\rho_1}-\frac{L_{b 2}}{\rho_2}+\frac{L_{b 3}}{\rho_3}\right) \zeta_3,
\end{equation}
\begin{equation}\label{Eq9}
R_{26}\left(\zeta_1\right)=\frac{L_{b 1}}{\rho_1}, \quad R_{26}\left(\zeta_2\right)=\frac{L_{b 1}}{\rho_1}-\frac{L_{b 2}}{\rho_2}, \quad R_{26}\left(\zeta_3\right)=\frac{L_{b 1}}{\rho_1}-\frac{L_{b 2}}{\rho_2}+\frac{L_{b 3}}{\rho_3},
\end{equation}
here $\zeta_j~(j = 1, 2, 3)$ are the local path-length coordinates in the three drift sectiions.

The above matrix elements are evaluated from the very beginning of the four-bend chicane. Since the CSR fields are generated inside the bending magnets, analyzing their dynamical impact on the beam also requires knowledge of the relative transfer matrices evaluated from each bending magnet to the lattice exit, i.e.,
\begin{equation}\label{Eq10}
\begin{aligned}
R_{16}\left(s \rightarrow s_f\right) &= -R_{16}(s)-\left[R_{12}\left(s_f\right)-R_{12}(s)\right] R_{26}(s), \\
R_{26}\left(s \rightarrow s_f\right) &= -R_{26}(s),
\end{aligned}
\end{equation}
where $R_{12}(s) = s$ and $R_{16}(s) = R_{16}(0 \to s), R_{26}(s) = R_{26}(0 \to s)$ are given above. The above expressions will be used to estimate the CSR-induced emittance growth. Moreover, we have the relative longitudinal momentum compaction function evaluated from some bending magnet to another 
\begin{equation}\label{Eq11}
\begin{aligned}
R_{56}\left(s_1 \rightarrow s_f\right) & =R_{56}, \\
R_{56}\left(s_2 \rightarrow s_f\right) & =R_{56}-R_{56}\left(s_2\right)=R_{56}-\frac{L_{b 1} L_{d 1} s_2}{\rho_1 \rho_2} \equiv R_{56}\left(1-\frac{s_2}{\overline{L_{b 2}}}\right), \\
R_{56}\left(s_3 \rightarrow s_f\right) & =R_{56}-R_{56}\left(s_3\right)=\left[R_{56}\left(s_{2 f}\right)-R_{56}\right]\left(\frac{s_3}{L_{b 3}}-1\right),
\end{aligned}
\end{equation}
here we define the shorthand notation
\begin{equation}\label{Eq12}
\overline{L_{b 2}} \equiv \frac{\rho_1 \rho_2 R_{56}}{L_{b1} L_{d1}}.
\end{equation}

In addition, we have
\begin{equation}\label{Eq13}
\begin{aligned}
R_{56}\left(s_1 \rightarrow s_2\right) & =\frac{L_{d 1} s_2}{\rho_1 \rho_2}\left(L_{b 1}-s_1\right) =\frac{R_{56}}{\overline{L_{b 2}}} s_2\left(1-\frac{s_1}{L_{b 1}}\right), \\
R_{56}\left(s_1 \rightarrow s_3\right) & =R_{56}\left(s_3\right)-\frac{R_{52}\left(s_3\right)}{\rho_1} s_1, \\
R_{56}\left(s_2 \rightarrow s_3\right) & =\frac{L_{d 2} s_3\left(L_{b 2}-s_2\right)}{\rho_2 \rho_3}.
\end{aligned}
\end{equation}
The above expressions will be used to estimate the CSR-induced microbunching gain. Here we remind that, as also mentioned at the beginning of this section, when considering the S-shaped chicane shown in Fig.~\ref{Fig1}, here $\rho_1, \rho_2, \rho_3, \rho_4$ are all assumed positive numbers. For a C-shaped chicane, since the polarity of the third and fourth magnets is opposite to that of the S-shaped chicane, the above corresponding pure optics functions for the C-shaped chicane can be obtained simply by taking $\rho_3 < 0, \rho_4 < 0$.

To describe the evolution of transverse beam phase space ellipse, the Courant-Snyder or Twiss parameters are used. Assuming $\beta_{x0}, \alpha_{x0}, \gamma_{x0}$ at the entrance of the four-bend chicane, we have at the exit of the lattice
\begin{equation}\label{Eq14}
\begin{aligned}
\beta_{xf} &= \beta_{x0}-2 \alpha_{x0} L_T+\gamma_{x0} L_T^2, \\
\alpha_{xf} &= \alpha_{x0}-\gamma_{x0} L_T.
\end{aligned}
\end{equation}

An energy-chirped beam will go through bunch compression (or decompression) in a longitudinal dispersive section with nonzero $R_{56}$. The performance of bunch compression is usually quantified by the linear compression factor, defined as
\begin{equation}\label{Eq15}
C \equiv \frac{\sigma_{z, s_0}}{\sigma_{z, s_f}}=\frac{\sigma_{z 0}}{\sqrt{R_{56}^2 \sigma_{\delta 0}^2+\left(R_{56} h \sigma_{z 0}+\sigma_{z 0}\right)^2}} \approx \frac{\sigma_{z 0}}{\left|R_{56} h \sigma_{z 0}+\sigma_{z 0}\right|}=\frac{1}{\left|1+h R_{56}\right|},
\end{equation}
where the chirp parameter is defined as $h = \partial \delta/\partial z$. The approximation holds when the beam is cold, i.e., the initial uncorrelated energy spread $\sigma_{\delta 0}$ much smaller than the correlated energy spread $h\sigma_{z0}$.

\subsection{Steady-state 1-D CSR model}\label{SecIIB}

The previous section summarized the pure-optics results needed for the subsequent analysis. In order to derive the CSR-induced bunching factor and emittance growth, this section will introduce the expressions of the CSR impedance and the induced energy spread. The steady-state CSR impedance per unit length can be written as~\cite{Ref02}
\begin{equation}\label{Eq16}
Z_\mathrm{CSR}^{\text{ss}}(k ; s)=-i A \times \frac{Z_0}{4 \pi} \frac{k(s)^{\frac{1}{3}}}{|\rho(s)|^{\frac{2}{3}}}, \text { with } A=1.63 i-0.94,
\end{equation}
where $k = 2\pi/\lambda$ is the wavenumber, $Z_0 \approx 377~\Omega$ is the free-space impedance, and $\rho$ is the radius of curvature of the corresponding bending magnet. Simulation results have shown that in addition to the steady-state CSR-induced amplification in the bunching factor and emittance growth, the impact of the transient CSR fields, particularly the exit transient~\cite{Ref03}, cannot be neglected~\cite{Ref32}. The complete expressions for the exit-transient CSR and its impedance~\cite{Ref14} could be complicated, and likely do not have an easy-to-use expression. Here, we adopt a simplified model that approximates the exit-transient CSR impedance as the edge radiation~\cite{Ref33}, given by
\begin{equation}\label{Eq17}
Z_\mathrm{CSR}^{\text {drift}}(k ; s) \approx \frac{Z_0}{4 \pi}\left\{\begin{array}{ll}
0, & \text { if } s^*<\rho^{\frac{2}{3}} \lambda(s)^{\frac{1}{3}} \\
\frac{2}{s^*}, & \text { if } \rho^{\frac{2}{3}} \lambda(s)^{\frac{1}{3}} \le s^* \le \frac{\lambda(s) \gamma^2}{2 \pi} \\
\frac{2 k(s)}{\gamma^2}, & \text { if } s^* \ge \frac{\lambda(s) \gamma^2}{2 \pi}
\end{array}\right.,
\end{equation}
where $\rho$ is bending radius, $s^*$ is the downstream distance measured from the exit of a bending magnet, $\gamma$ is the relativistic factor, and $\lambda$ is the wavelength. In Eq. (\ref{Eq17}), the term $\rho^{2/3}\lambda^{1/3} = \ell_{Bf}$ is characteristic of the formulation length of synchrotron radiation inside a bending magnet, while the term $\lambda \gamma^2 = \ell_{Df}$ can be roughly considered as the formation length outside a magnet on the drift section. The expression is valid only when $\ell_{Bf} < \ell_{Df}$, which is well satisfied in our case. For very low energy, Eq. (\ref{Eq17}) may become invalid.

Here we remind that the complete set of CSR fields consists of four parts, corresponding to Case A to Case D in Ref.~\cite{Ref03}. In the following derivation of the CSR-induced microbunching instability, we only consider the steady-state CSR impedance and the exit-transient CSR impedance, which we believe cover the two main contributing physical effects.

The CSR on the beam in the transverse phase space is manifested in the increase of the projected emittance. Although there has been extensive work on the impact of CSR on the beam, most of it assumes a steady-state CSR, i.e., the effect of Eq. (\ref{Eq16}) on the beam. In recent years, Khan et al.~\cite{Ref34} has discussed in depth the change in the beam energy spread due to the transient CSR fields as it passes through a bending magnet, providing a semi-analytical expression. We will carry over from their work. To facilitate the subsequent derivations in this paper, here we summarize the main results of Ref.~\cite{Ref34}. When the beam passes through a single bending magnet, the increase in energy spread due to the entrance transient and steady-state CSR fields can be expressed as
\begin{equation}\label{Eq18}
\begin{aligned}
&\Delta \sigma_{\delta,\text{CSR,in-bend}} \approx \frac{N r_e L_B}{\gamma} \times \\
&\begin{cases}0.029 \frac{\left(2 \sigma_{z i}-\sigma_{z f}\right)^2 \theta_T^2}{\sigma_{z i}^4}, & \text { if } \theta_T>\phi_m \\ \left[ 0.029 \frac{\left(2 \sigma_{z i}-\sigma_{z f}\right)^2 \theta_T^2}{\sigma_{z i}^4}+\frac{0.738}{\rho^{2 / 3}\left(\sigma_{z i}-\sigma_{z f}\right)}\left(\frac{1}{\sigma_{z f}^{1 / 3}}-\frac{1}{\left(\sigma_{z i}-\left(\frac{\sigma_{z i}-\sigma_{z f}}{\phi_{m}}\right) \theta_T\right)^{1 / 3}}\right)\right] & \text { if } \theta_T<\phi_m\end{cases}
\end{aligned}
\end{equation}
with $N$ being the number of particles in the beam, $r_e \approx 2.818 \times 10^{-15}~\mathrm{m}$ the classical radius of electron, $L_B = \rho \phi_m$ the magnet length, with $\rho$ and $\phi_m$ the bending radius and angle. Here, $\sigma_{zi}$ and $\sigma_{zf}$ refer to the bunch length at the entrance and exit of a bending magnet, respectively. In Eq. (\ref{Eq18}), the transition from entrance transient state to steady state is quantified by the transition angle $\theta_T$, which can be determined by solving the following cubic equation
\begin{equation}\label{Eq19}
1.6 \frac{\rho}{24} \theta_T^3-\left(\frac{\sigma_{z f}-\sigma_{z i}}{\phi_m}\right) \theta_T-\sigma_{z i}=0.
\end{equation}
Having solved this cubic equation to obtain the $\theta_T$, we substitute into Eq. (\ref{Eq18}) to calculate the increase in energy spread due to a single bending magnet. When the beam exits the bending magnet, it will be affected by the exit-transient CSR field downstream, leading to an additional increase in the energy spread given by
\begin{equation}\label{Eq20}
\Delta \sigma_{\delta,\text{CSR,out-bend}} \approx 0.22\left(1-e^{\Pi \rho \phi_m^3}\right) \frac{N r_e}{\gamma \sigma_z} \ln \left(1+\chi \frac{2 \zeta}{L_B}\right),
\end{equation}
where $\sigma_z$ is the bunch length at the exit of the (upstream) bending magnet and $\zeta$ is the downstream drift distance. The two empirical functions are
\begin{equation}\label{Eq21}
\Pi \approx 0.132 \zeta^{1 / 2} \sigma_z^{-3 / 2}, \quad \chi  \approx 0.525\left(\frac{\rho \phi_m^3}{\sigma_z}\right)^{1 / 3}.
\end{equation}

So far, we have collected the necessary ingredients. The summarized CSR impedances above will be used in deriving the MBI gain, while the CSR-induced energy spread increase will be utilized in deriving the projected emittance of the beam in an arbitrary achromatic four-bend chicane.

\subsection{CSR-induced microbunching amplification gain}\label{SecIIC}

Before deriving the CSR-induced microbunching amplification gain, we first define the longitudinal bunching factor of the beam as the Fourier transform of the (perturbed) phase space distribution function
\begin{equation}\label{Eq22}
b(k ; s)=\frac{1}{N} \int d \mathbf{X} e^{-i k z} f(\mathbf{X} ; s),
\end{equation}
where the phase space distribution $f = f_0 + f_1$ has a small perturbation on top of the smooth background, $|f_1| \ll f_0$. When the smooth background distribution is uniform in $z$ and the characteristic length of perturbation is much shorter than the bunch length, it does not contribute to the resulting bunching factor, i.e., the dynamics of $b(k;s)$ is determined by the perturbed term $f_1$. Here, $k = 2\pi/\lambda$ is the modulation wavenumber, and $\mathbf{X} = (x, x', z, \delta)^T$ represents the 4-D phase space coordinates introduced earlier. Since the energy changes induced by the CSR fields at different positions within the beam are correlated as the beam propagates through downstream non-zero dispersion sections $R_{56}$, the energy modulation may translate into density modulation, leading to collective effects that amplify the bunching factor of the beam. Omitting the derivation details, the established theory~\cite{Ref08,Ref09} indicates that the evolution of this bunching factor along the transport satisfies the following integral equation
\begin{equation}\label{Eq23}
b[k(s) ; s]=b_0[k(s) ; s]+\int_0^s d \tau K(\tau, s) b[k(\tau) ; \tau],
\end{equation}
where $b_0[k(s) ; s]$ is the bunching factor in the absence of collective effects (i.e., the pure-optics term), and the term contributing to the dynamical evolution, the kernel function, can be expressed as
\begin{equation}\label{Eq24}
\begin{aligned}
K(\tau, s) & =i k(s) R_{56}(\tau \rightarrow s) \frac{4 \pi}{Z_0} \frac{I_b(\tau) Z_{\|}[k(\tau) ; \tau]}{\gamma I_A} e^{-\frac{k_0^2}{2} U^2(s, \tau) \sigma_{\delta 0}^2} \\
& \times \exp \left[-\frac{k_0^2 \epsilon_{x 0} \beta_{x 0}}{2}\left(V(s, \tau)-\frac{\alpha_{x 0}}{\beta_{x 0}} W(s, \tau)\right)^2-\frac{k_0^2 \epsilon_{x 0}}{2 \beta_{x 0}} W^2(s, \tau)\right],
\end{aligned}
\end{equation}
where $I_b(\tau)$ is the bunch current evaluated at the location $\tau$, $I_A=\frac{e c}{r_e} \approx 17045 ~\mathrm{A}$ is the Alfven current, $k_0 = k(s_0)$ is the initial wavenumber, $\epsilon_{x0}$ is the initial geometric beam emittance, $\sigma_{\delta 0}$ is the initial energy spread, and $\beta_{x0}, \alpha_{x0}$ are the Courant-Snyder parameters of the beam. In the exponent of Eq. (\ref{Eq24}), 
\begin{equation}\label{Eq25}
\begin{aligned}
U(s, \tau) & =C(s) R_{56}(s)-C(\tau) R_{56}(\tau), \\
V(s, \tau) & =C(s) R_{51}(s)-C(\tau) R_{51}(\tau), \\
W(s, \tau) & =C(s) R_{52}(s)-C(\tau) R_{52}(\tau).
\end{aligned}
\end{equation}
The above exponential terms in the kernel characterize the warm beam effects of the bunch, including the Landau damping or phase space smearing effect due to the transverse emittance and the longitudinal energy spread. 

Equation (\ref{Eq23}) is applicable to general linear transport lines. Typical methods to solve Eq. (\ref{Eq23}) are through numerical integration or matrix inversion method, such as the semi-analytical Vlasov solver we recently developed~\cite{Ref11,Ref12,Ref15}. If the number of bending magnets is not large, this equation can also be solved analytically through iterative methods, such as the gain formula for the symmetric C-shaped chicane derived by Huang and Kim~\cite{Ref09}. Now, we will use the analytical iterative approach to derive the MBI gain formula for the achromatic four-bend chicane shown in Fig.~\ref{Fig1}. Due to the addition of one more bending magnet, we will iterate up to the third order, and in Sec.~\ref{SecIII}, we will compare the results of this formula with those from a semi-analytical Vlasov solver for two example lattices. Although the semi-analytical Vlasov calculation is much faster than time-consuming particle tracking simulations, we want to emphasize that since we have obtained an analytical formula, the computation speed will be even faster than the Vlasov calculation, making it more suitable for comprehensive multi-parameter scans and optimization of the achromatic four-bend chicane lattice design.

According to the iterative method, Eq. (\ref{Eq23}) can be split into the following integrals
\begin{equation}\label{Eq26}
\begin{aligned}
b[k(s_f) ; s_f] &\approx b_0[k(s_f) ; s_f] \\
&+\int_0^{L_{b 1}} d s_1 K\left(s_1, s_f\right) b_0\left[k\left(s_1\right) ; s_1\right]\\
&+\int_0^{L_{b 2}} d s_2 K\left(s_2, s_f\right) b_0\left[k\left(s_2\right) ; s_2\right]\\
&+\int_0^{L_{b 3}} d s_3 K\left(s_3, s_f\right) b_0\left[k\left(s_3\right) ; s_3\right] \\
&+\int_0^{L_{b 2}} d s_2 K\left(s_2, s_f\right) \int_0^{L_{b 1}} d s_1 K\left(s_1, s_2\right) b_0\left[k\left(s_1\right) ; s_1\right] \\
&+\int_0^{L_{b 3}} d s_3 K\left(s_3, s_f\right) \int_0^{L_{b 1}} d s_1 K\left(s_1, s_3\right) b_0\left[k\left(s_1\right) ; s_1\right] \\
& +\int_0^{L_{b 3}} d s_3 K\left(s_3, s_f\right) \int_0^{L_{b 2}} d s_2 K\left(s_2, s_3\right) b_0\left[k\left(s_2\right) ; s_2\right] \\
&+\int_0^{L_{b 3}} d s_3 K\left(s_3, s_f\right) \int_0^{L_{b 2}} d s_2 K\left(s_2, s_3\right) \int_0^{L_{b 1}} d s_1 K\left(s_1, s_2\right) b_0\left[k\left(s_1\right) ; s\right] \\
&+  \int_0^{L_{d 1}} d \zeta_1 K\left(\zeta_1, s_f\right) b_0\left[k\left(\zeta_1\right) ; \zeta_1\right]\\
&+ \int_0^{L_{d2}} d \zeta_2 K\left(\zeta_2, s_f\right) b_0\left[k\left(\zeta_2\right) ; \zeta_2\right] \\
&+ \int_0^{L_{d 3}} d \zeta_3 K\left(\zeta_3, s_f\right) b_0\left[k\left(\zeta_3\right) ; \zeta_3\right],
\end{aligned}
\end{equation}
where the first term on the right hand side refers to the pure-optics term, the second to the fourth terms correspond to the first iteration, the fifth to the seventh terms to the second iteration and the eighth term to the third iteration. It is found that the third-iteration term is negligibly small in our cases. In contrast to the contributions from steady-state CSR-induced microbunching, the last three integrals correspond to the contribution from the exit-transient CSR fields. We define the microbunching gain at the exit of a lattice as
\begin{equation}\label{Eq27}
G_f \equiv \left| \frac{b\left(k_f ; s_f\right) }{ b_0\left(k_0 ; s_0\right)}\right|.
\end{equation}

Although Eq. (\ref{Eq26}) appears cumbersome, we will proceed by evaluating each integral individually and then summing the obtained results to form the final gain formula. For clarity and to avoid tedious calculations, here we only summarize the main results, while the explicit forms of the sub-terms can be found in ~\ref{AppA}.
\scriptsize
\begin{equation}\label{Eq28}
\begin{aligned}
G_f \approx \left|\begin{array}{l}
\left.\exp \left[-\frac{\bar{\sigma}_\delta^2}{2\left(1+{hR}_{56}\right)^2}\right]+{A} \overline{{I}}_{{f} 1}\left[{F}_0+\frac{1-{e}^{-\bar{\sigma}_{{x1}}^2}}{2 \bar{\sigma}_{{x}1}^2}\right] \times \exp \left[-\frac{\bar{\sigma}_\delta^2}{2\left(1+{h} {R}_{56}\right)^2}\right]+{A} \overline{{I}}_{f2} ~{F}_1+{A} \overline{{I}}_{{f} 3} {~F}_2 \right. \\
\left. + {A}^2 \overline{{I}}_{{f} 1} \overline{{I}}_{{f} 2} {~F}_0 {F}_3+ {A}^2 \overline{{I}}_{{f} 1} \overline{{I}}_{{f} 3}\left[{~F}_4 \times \left({~F}_0+\frac{1-{e}^{-\bar{\sigma}_{{x} 1}^2}}{2 \bar{\sigma}_{{x} 1}^2}\right) - {F}_5 \frac{1-{e}^{-\bar{\sigma}_{{x}}^2}}{2 \bar{\sigma}_{{x} 1}^2}\right]+{A}^2 \overline{{I}}_{{f} 2} \overline{{I}}_{{f} 3} {~F}_6 {~F}_7 \right. \\
\left.+{A}^3 \overline{I}_{f 1} \overline{{I}}_{f 2} \overline{I}_{f 3} ~F_0 ~F_7 ~F_8 \right. \\
\left.+D_1+D_2+D_3 \right.
\end{array}\right|
\end{aligned}
\end{equation}
\normalsize
with $A = 1.63i – 0.94$. The first term in Eq. (\ref{Eq28}), independent of the scaled bunch current $\bar{I}_f$ corresponds to the contribution from pure optics, the terms linearly proportional to the current $\bar{I}_{fj}~(j=1,2,3)$ are the first-order amplification terms (corresponding to the first-iteration result), the terms proportional to $\bar{I}_{fi}\bar{I}_{fj}$ are the second-order amplification terms (corresponding to the second-iteration result), and so on. When this MBI gain is large, say 100, then the beam is subject to the microbunching instability, meaning that the lattice will amplify any initial density perturbation of the bunch entering the lattice by a factor of approximately 100 (if not yet saturated). Generally, one aims to design the lattice to avoid excessively large gains. The explicit forms of the accompanying $F_k$ functions can be found in \ref{AppA}. It is worth mentioning that $D_1$, $D_2$, and $D_3$ here correspond to the impact of the exit-transient CSR, whose contributions to the final gain $G_f$ cannot typically be neglected. Regarding the applicability of this formula, we remind that for the S-shaped chicane [see Fig.~\ref{Fig1}], $\rho_i > 0$ with $i=1,2,3,4$. For the C-shaped chicane, then $\rho_{1,2} > 0, \rho_{3,4} < 0$. 

It is straightforward to check that for the case of a C-shaped symmetric chicane with three bending magnets, Eq. (\ref{Eq28}) can reduce to the well-known formula in Ref.~\cite{Ref09}. Alternatively, one can also view this result of our derived four-bend S-shape chicane as physically equivalent to the symmetric C-shaped chicane as a special geometric configuration: a C-shaped chicane can be seen as an S-shape, when only looking at the first three bending magnets in Fig.~\ref{Fig1} by setting the bending radius of the fourth magnet tending towards infinity (i.e., no bending action of the fourth magnet).

\subsection{CSR-induced emittance growth}\label{SecIID}

In addition to amplification of the bunching factor of the beam, the CSR-induced energy deviations may also lead to an increase in the transverse projected emittance through the non-zero dispersion terms $R_{16}$ and $R_{26}$. According to Refs.~\cite{Ref04,Ref35}, the increase in the transverse projected emittance due to the coherent interactions can be expressed as
\begin{equation}\label{Eq29}
\epsilon_{xf}^2=\epsilon_{x0}^2+ \frac{\epsilon_{x0}}{\beta_{xf}}\left[\left\langle\Delta x^2\right\rangle_\mathrm{coh}+\left(\alpha_{xf}\left\langle\Delta x^2\right\rangle_\mathrm{coh}^{1 / 2}+\beta_{xf}\left\langle\Delta x^{\prime 2}\right\rangle_\mathrm{coh}^{1 / 2}\right)^2\right],
\end{equation}
where the $\beta_{xf}, \alpha_{xf}$ are the Courant-Snyder parameters at the lattice exit. The total squared changes of the transverse beam size and divergence can be obtained by integrating the CSR-induced effects along the transport line~\cite{Ref04,Ref35}
\begin{equation}\label{Eq30}
\begin{aligned}
\left\langle\Delta x^2\right\rangle_\mathrm{coh} &= \left(\int_{s_0}^{s_f} R_{16}\left(s \rightarrow s_f\right) \frac{\partial \sigma_{\delta,\text{CSR}}}{\partial s} d s\right)^2,\\
\left\langle\Delta x^{\prime 2}\right\rangle_\mathrm{coh} &= \left(\int_{s_0}^{s_f} R_{26}\left(s \rightarrow s_f\right) \frac{\partial \sigma_{\delta,\text{CSR}}}{\partial s} d s\right)^2,
\end{aligned}
\end{equation}
where the increase in the CSR-induced beam energy spread has been summarized in Sec.~\ref{SecIIB}. Now we intend to split the integration into several individual segments, including those within the four bending magnets and three drift sections. In total there are seven integrals for Eqs. (\ref{Eq30}) for the beam size and divergence expressions, respectively. By substituting the ingredients from Sec~\ref{SecIIA} and Sec~\ref{SecIIB} into the above integrals and performing tedious but straightforward calculations, we can obtain the formula for the final emittance growth. For the change in energy spread per unit length due to CSR in the integral expressions, we make an approximation
\begin{equation}\label{Eq31}
\frac{\partial \sigma_{\delta, \mathrm{CSR}}}{\partial s} \approx \frac{\Delta \sigma_{\delta,\text{CSR}}}{L_{b} ~\text{or}~L_d},
\end{equation}
where, depending on the situation, the denominator can be $L_{b}$ (the length of a bending magnet) or $L_d$ (the length of a drift section). This approximation simplifies the calculations. Similarly, for clarity and to avoid tedious calculations, here we only summarize the main results, while the explicit forms of the sub-terms can be found in \ref{AppB}.
\begin{equation}\label{Eq32}
\begin{aligned}
\left\langle\Delta x^2\right\rangle_{\text {coh }}&=\left(g_1+g_2+g_3+g_4+g_5+g_6+g_7\right)^2, \\
\left\langle\Delta x^{\prime 2}\right\rangle_{\text {coh }}&=\left(g_8+g_9+g_{10}+g_{11}+g_{12}+g_{13}+g_{14}\right)^2.
\end{aligned}
\end{equation}
Here we note that, in deriving the CSR-induced emittance growth, for simplicity we have used the first-order approximations for $R_{16}$ and $R_{26}$, neglecting the effects of higher-order terms.

Up to this point, we have presented the main results of this paper, namely the analytical formulas for the CSR-induced MBI gain [Eq. (\ref{Eq28})] and the formula for emittance growth [Eqs. (\ref{Eq29}) and (\ref{Eq32})] in an arbitrary achromatic four-bend chicane. Since they are analytical formulas, the computational efficiency will be greatly improved, making them suitable for quick evaluation of CSR effects during the preliminary design stage of the chicane lattice. Furthermore, with the provided expressions, the functional dependences on the relevant lattice and electron beam parameters are explicitly given, facilitating comprehensive multi-parameter scans for optimization. By comparing to the formulas already in the existing literature, we emphasize that the formulas derived in this paper include both the steady-state and transient effects of CSR.

\section{Numerical examples}\label{SecIII}

As an application of the analytical formulas derived in the previous section, in this section we will consider two examples of achromatic four-bend chicanes: a widely used symmetric C-shaped chicane and a non-symmetric S-shaped chicane. It is worth mentioning that recently, our co-authors, based on the point-kick model in Ref.~\cite{Ref27}, proposed an innovative non-symmetric, four-bend S-shaped chicane bunch compressor design~\cite{Ref36,Ref37}. In addition to suppressing emittance growth, this design can also effectively mitigate MBI. For a more detailed discussion on the design rationale of the non-symmetric S-shaped chicane, we refer the interested readers to Refs.~\cite{Ref36,Ref37}.

\subsection{Symmetric C-shape chicane}\label{SecIIIA}

As the first example, we consider the symmetric C-shaped chicane, with the schematic diagram shown in Fig.~\ref{Fig1}, where the third bending magnet is represented by the blue block. Table 1 summarizes the initial beam and lattice design parameters, which are roughly based on the preliminary design parameters for the LCLS-II BC2 chicane a few years ago~\cite{Ref34}.

\begin{table}\label{TableI}
\centering
\caption{Initial beam and lattice parameters for the example of the symmetric C-shape chicane}
\begin{tabular}{lcl}
\hline\hline
Name                      & Value  & Unit \\ \hline
Electron energy           & 5.01   & GeV  \\
Beam energy spread        & $1 \times 10^{-6}$     &      \\
Bunch charge              & 100    & pC   \\
Initial bunch length      & 54     & $\mu$m   \\
Initial bunch current     & 237    & A    \\
Normalized beam emittance & 0.41   & $\mu$m   \\
Initial beta function     & 120~(may vary)    & m    \\
Initial alpha function    & 5.5~(may vary)    &      \\
Initial energy chirp      & $-39.52$ &  m$^{-1}$    \\
Total compression factor  & 41.8   &      \\
Magnet length $(L_{b1},L_{b2},L_{b3},L_{b4})$            &  $(0.55, 0.55, 0.55, 0.55)$  & m    \\
Bending radius $(\rho_{1},\rho_{2},\rho_{3},\rho_{4})$           &  $(15.84, 15.84, 15.84, 15.84)$  & m    \\
Drift length $(L_{d1},L_{d2},L_{d3})$             &  $(9.87, 1.09, 9.87)$ & m    \\ \hline\hline
\end{tabular}
\end{table}

\begin{figure}[htbp]
\centering
\includegraphics[width=12cm]{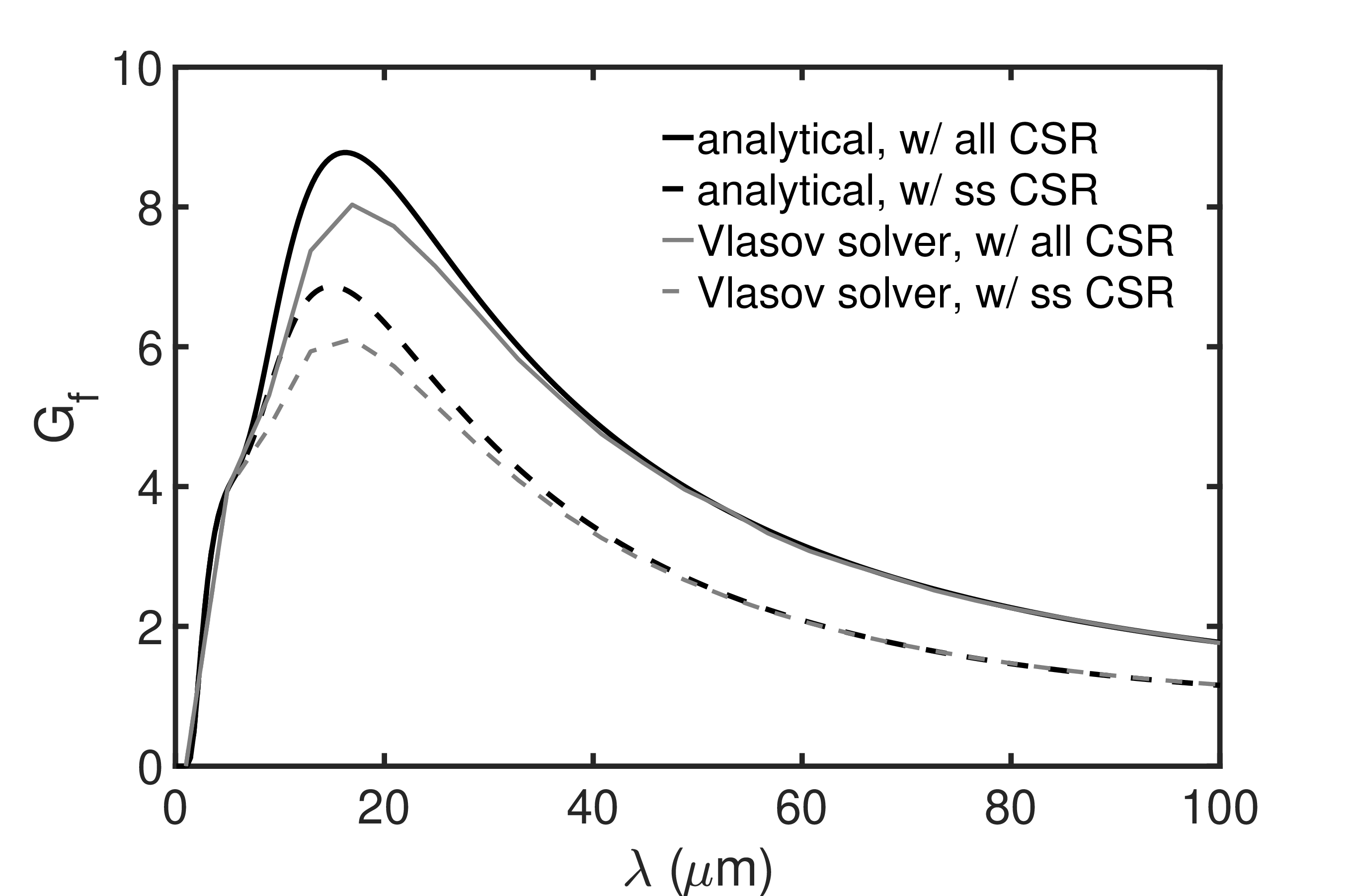}
\caption{\textmd{Typical microbunching gain spectra of a symmetric C-shape chicane. The black lines are obtained from our derived analytical formula. The gray lines are from semi-analytical calculations by Vlasov solver. The solid lines contain all CSR (including steady-state and transient CSR) impedances, while the dashed lines only include the steady-state CSR.}}
\label{Fig2}
\end{figure}

Figure~\ref{Fig2} shows the calculated results of the CSR-induced microbunching instability (MBI) gain based on the analytical formula derived in the previous section and the semi-analytical Vlasov calculations. For those short modulation wavelength components (i.e., large $k$), the obvious phase space smearing (or Landau damping) caused by the non-zero $R_{51}, R_{52}, R_{56}$ makes it difficult to accumulate and develop microbunching amplification within the bunch. Although the longer modulation wavelengths are less susceptible to smearing caused by the non-zero transverse-longitudinal coupling, the smaller $k$ means a relatively weak CSR impedance [see Eq. (\ref{Eq16})]. Only those wavelength components that lack Landau damping but still retain a certain amount of CSR fields will continuously modulate the electron beam and form the MBI.

From Fig.~\ref{Fig2}, it can also be seen that the black solid line (all CSR) and the dashed line (only steady-state CSR) based on the analytical formula are in good agreement with the corresponding gray solid and dashed lines based on the semi-analytical Vlasov solver. We believe that the small difference (within 10\%) is due to both the the analytical formula and the numerical calculation. The numerical calculation of the integral equation [Eq. (\ref{Eq23})] involves discretizing the $\tau$-integral, and the finite number of meshes will inevitably produce some differences. In this calculation, the total length of the lattice is about 34 m, and taking 1000 meshes can already make the results converge. Another possible cause for the difference is due to the additional approximations made in order to obtain the analytical formula to simplify the integration (see \ref{AppA}). For the case with exit-transient CSR, when we substitute the drift CSR impedance [Eq. (\ref{Eq17})] into the integral equation, we only consider the first-order iteration for simplicity. In addition, in using the analytical formula, we found that the value of the final gain may depend sensitively on the value of $R_{56}$ of the entire transport line, especially for the large compression factor. Mathematically, this is because $R_{56}$ appears frequently in the numerator or denominator of the integral of the analytical formula, especially in the denominator. The small difference in the value of $R_{56}$ may have a certain impact on the final gain. Physically, a slightly different $R_{56}$ value means that the bunch compression process is slightly different, which involves slightly different dynamical evolution in the bunch compression, leading to different results of the MBI gain. Overall, the analytical results are in good agreement with the results of the semi-analytical Vlasov calculations.

\begin{figure}[htbp]
\centering
\includegraphics[width=14cm]{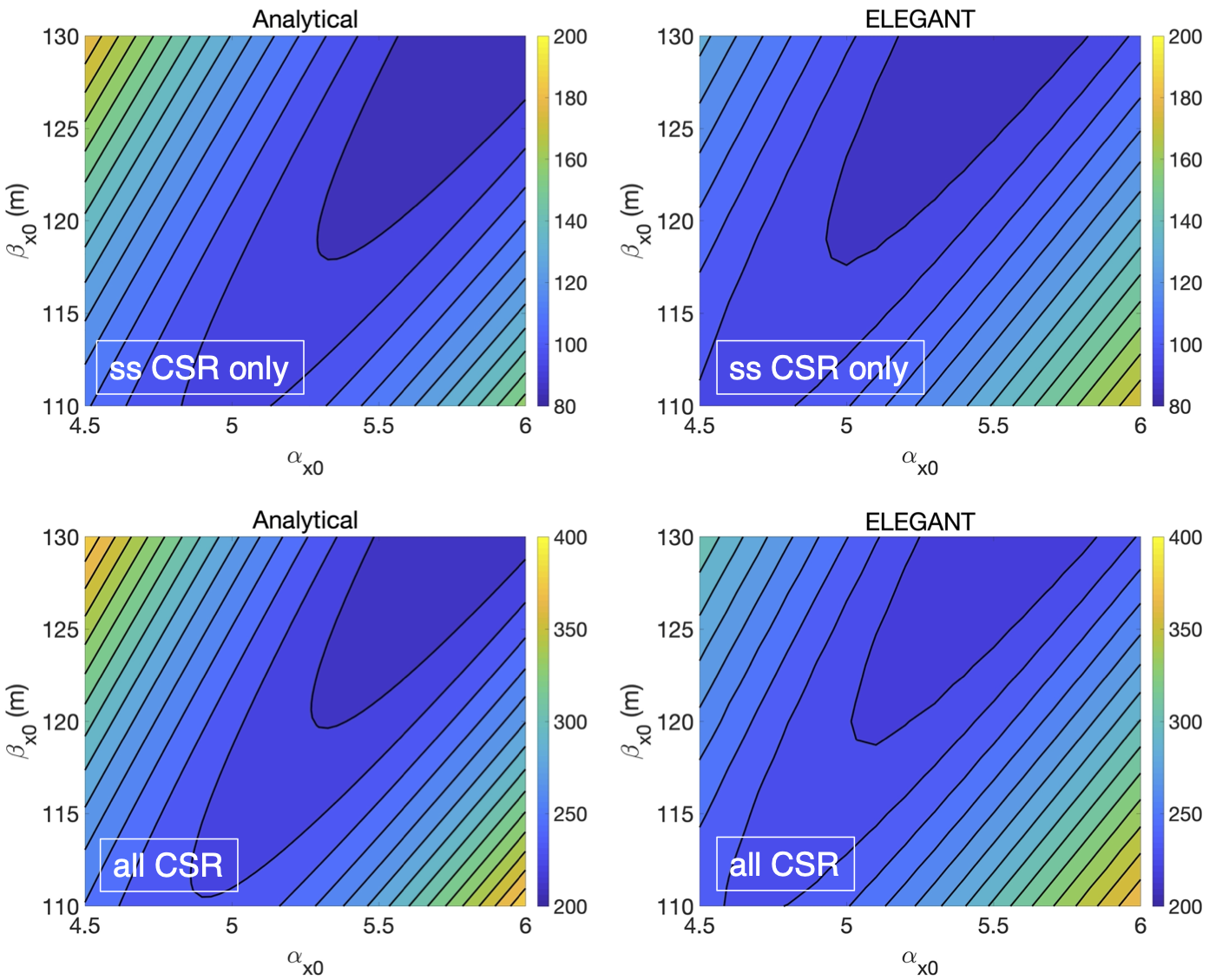}
\caption{\textmd{Relative emittance growth (in \%) vs. initial Courant-Snyder parameters. The plots on the top row include only the steady-state CSR; those on the bottom row contain both steady-state and transient CSR (i.e., all CSR). The left plots are obtained from analytical expressions; the results on the right plots are collected from ELEGANT particle tracking simulations.}}
\label{Fig3}
\end{figure}

Figure~\ref{Fig3} shows the calculated results of the analytical formula for the emittance growth derived in the previous section and the results of the ELEGANT particle tracking simulation~\cite{Ref38,Ref39}. In Fig.~\ref{Fig3}, we scan a series of initial Courant-Snyder parameters $(\beta_{x0}, \alpha_{x0})$ of the beam at the entrance of the lattice, and then record the relative emittance growth at the exit of the lattice. The left column shows the results obtained by scanning the parameters according to the analytical formula, and the right column shows the results obtained by ELEGANT. The upper row only considers the steady-state CSR induced emittance growth, while the lower row includes the effects of both steady-state and the transient CSR. Although there are some differences, especially some offsets for the initial $\alpha_{x0}$, the results obtained by the analytical formula are in good agreement with the results by the time-consuming particle tracking simulations.

When running ELEGANT, $10^6$ simulation particles were used to represent a bunch in order to obtain converged results. In the bending magnet, when using the element \texttt{CSRCSBEND}, the propagation is cut into 20 parts (N\_KICKS), and the histogram required for each calculation of the CSR wake is set to 1000 parts (N\_BINS). The numerical noise is appropriately filtered using the built-in function of ELEGANT. In order to better compare with the analytical model, we turned off the incoherent synchrotron radiation (ISR) effect in the bending magnet. In the drift section, we use the \texttt{CSRDRIFT} element, where the step length is set as 0.1 m, and the Stupakov-Emma model is adopted~\cite{Ref40}. We believe that these settings can make the calculation of the emittance growth converge~\cite{Ref39,Ref3902}. However, we also note that, according to our previous experience, if we plan to use ELEGANT to measure the MBI gain, these simulation settings will become more stringent~\cite{Ref41}. These settings, especially the number of simulation particles and number of bins, may need to be increased.

Our first example demonstrates the agreement between the MBI gain obtained using the analytical formula and that obtained using the semi-analytical Vlasov solver. This is important in that the derived formula includes both the steady-state and exit-transient CSR. Furthermore, we use the analytical formula derived for the emittance growth to compare the results with those obtained by the particle tracking simulation. Again, there is a satisfactory agreement in the case where both the steady-state and exit transient CSR are included.

Although our analysis in this paper is not limited to the symmetric C-shape chicanes, the consistency obtained by using such a classic example and comparing with the long-lasting numerical tool gives us confidence in the analytical formula. In the next subsection, we will apply these formulas to a second example, the asymmetric S-shape chicane.

\subsection{Non-symmetric S-shape chicane}\label{SecIIIB}

As the second example, we consider a non-symmetric S-shape chicane, as shown in Fig.~\ref{Fig1}, where the third bending magnet is indicated by the red block. Table 2 summarizes the initial bunch and lattice design parameters, which are roughly based on an innovative four-bend S-shape chicane bunch compressor chicane, recently proposed by our co-authors~\cite{Ref36,Ref37}. Our analysis below shows that this design can effectively suppress emittance growth and mitigate MBI.

\begin{table}\label{TableII}
\centering
\caption{Initial beam and some lattice parameters for the example of the non-symmetric S-shape chicane}
\begin{tabular}{lcl}
\hline\hline
Name                      & Value  & Unit \\ \hline
Electron energy           & 3   & GeV  \\
Beam energy spread        & $2 \times 10^{-5}$     &      \\
Bunch charge              & 300~(may vary)    & pC   \\
Initial bunch length      & 100     & $\mu$m   \\
Initial bunch current     & 384    & A    \\
Normalized beam emittance & 0.9   & $\mu$m   \\
Initial beta function     & 66~(may vary)    & m    \\
Initial alpha function    & 4.8~(may vary)    &      \\
Initial energy chirp      & $-24$~(may vary) &  m$^{-1}$    \\
Total compression factor  & 10.2   &      \\
Magnet length $(L_{b1},L_{b2},L_{b3},L_{b4})$            &  $(0.5, 0.878, 0.48, 0.1)$  & m    \\
Bending radius $(\rho_{1},\rho_{2},\rho_{3},\rho_{4})$           &  $(7.782, 7.782, 7.782, 7.782)$  & m    \\
Drift length $(L_{d1},L_{d2},L_{d3})$             &  $(4.06, 7.46, 6.52)$ & m    \\ \hline\hline
\end{tabular}
\end{table}

\begin{figure}[htbp]
\centering
\includegraphics[width=12cm]{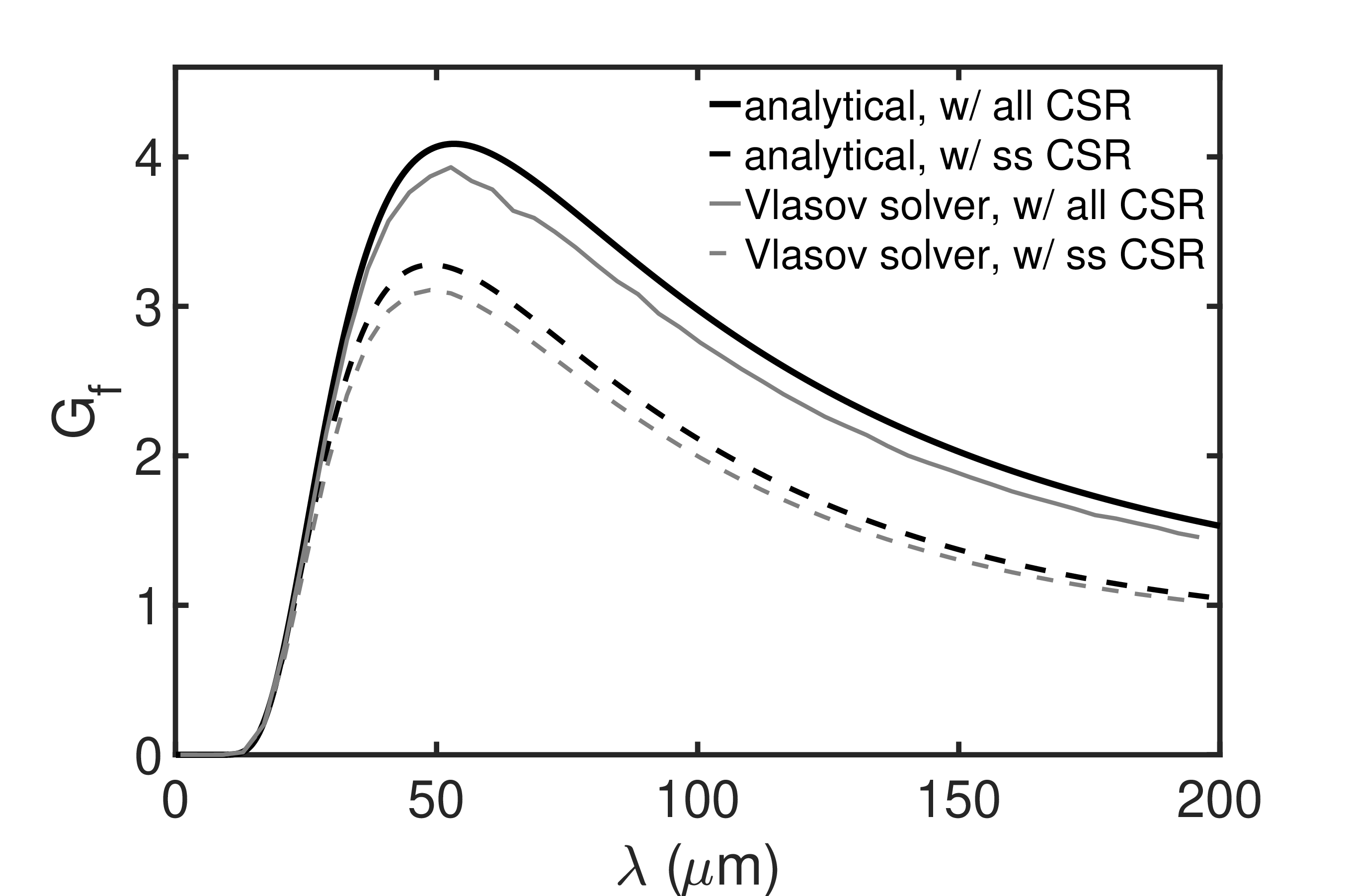}
\caption{\textmd{Typical microbunching gain spectra of a non-symmetric S-shape chicane. The black lines are obtained from our derived analytical solutions. The gray lines are from semi-analytical calculations by Vlasov solver. The solid lines contain all CSR (including steady-state and transient CSR) impedances, while the dashed lines only include the steady-state CSR.}}
\label{Fig4}
\end{figure}

We first consider the nominal case with the parameters in Table 2. Figure~\ref{Fig4} shows the CSR-induced MBI gain spectra. In this calculation, the total length of the chicane is about 20 m, and we still assume 1000 meshes to solve the integral equation. It can be seen that after the bunch is compressed from its peak current of 384 A to about 4 kA, the density modulation amplification of the bunch can be controlled below 4, even including both the steady-state and exit transient effects. This is the first time that we have applied the derived gain formula to the asymmetric S-shape chicane to obtain the result. The result is in good agreement with the result of the semi-analytical Vlasov calculation.

\begin{figure}[htbp]
\centering
\includegraphics[width=14cm]{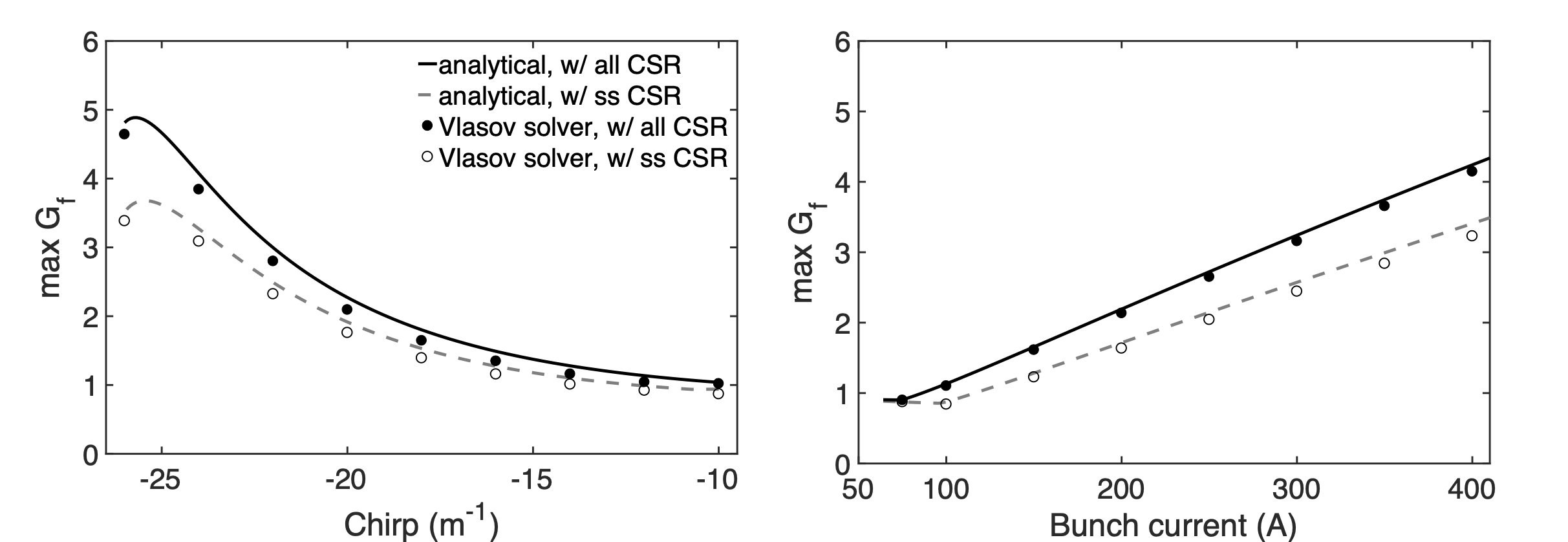}
\caption{\textmd{Maximal gain at the lattice exit as a function of the initial energy chirp and the initial bunch current for the non-symmetric S-shape chicane. The solid and dashed lines are calculated by the analytical formulas for the all CSR (including steady-state and transient CSR) and for the only steady-state CSR, respectively. The solid and empty circles represent the numerical results from Vlasov solver. For reference, the initial chirp $h = -10~\mathrm{m}^{-1}$ corresponds to a total of compression ratio $C = 1.6$, $h = -20~\mathrm{m}^{-1}$ to $C = 4$, and $h = -26~\mathrm{m}^{-1}$ to $C = 42$.}}
\label{Fig5}
\end{figure}

In addition to the greatly increased computational efficiency, the analytical formula can also be used to study the functional dependences of the gain on the parameters. In order to further understand how the MBI gain is related to the relevant physical quantities, we demonstrate two of the results, as shown in Fig.~\ref{Fig5}. The left figure of Fig.~\ref{Fig5} shows the change of the maximum gain when the initial energy chirp is different. Here we can see that when the energy chirp is small (in magnitude), the corresponding bunch compression factor is small, and the maximum gain is small. When the energy chirp is large (in magnitude), the corresponding bunch compression factor is large, and the maximum gain increases. It is worth noting that this trend is not linear, because when the bunch is compressed to a greater extent, although the peak current increases due to Liouville theorem in phase space, it is also accompanied by an increase in the effective energy spread, which will enhance Landau damping. The final gain value is obtained after the two effects counteract each other.

If the chirp is fixed and the initial current is changed, it can be seen from the right figure of Fig.~\ref{Fig5} that the maximum gain is roughly proportional to the initial current. Note that this linear relation is not a general result, but only that the present lattice parameters and bunch current range remain in the range where the maximum gain is linear with it. Although not shown here, when the initial current is increased to 1 kA, the maximum gain will no longer be proportional to the current. In fact, MBI in multi-bend transport line may exhibit multi-stage amplification, as revealed in Ref.~\cite{Ref42,Ref43}.

Here we would like to point out that it will become very difficult or almost impossible to systematically investigate these effects with ELEGANT. First, these effects all exist in the case of high bunch currents with strong collective effects. Therefore, in order to ensure the convergence and reliability of tracking simulation results, it is necessary to conduct convergence tests on many simulation parameters~\cite{Ref41}. Even using a cluster computer, a case generally takes hours to compute. In addition, the finite number of simulation particles always has a certain amount of numerical noise, which may lead to fluctuations in the results after post-processing the simulation data. Therefore, these functional dependences are easily buried within numerical fluctuations due to poor convergence. The analytical formula can avoid the influence of numerical artifact and requires almost no computation time; Fig.~\ref{Fig5} can be generated on a laptop in less than 30 seconds.

\begin{figure}[htbp]
\centering
\includegraphics[width=12cm]{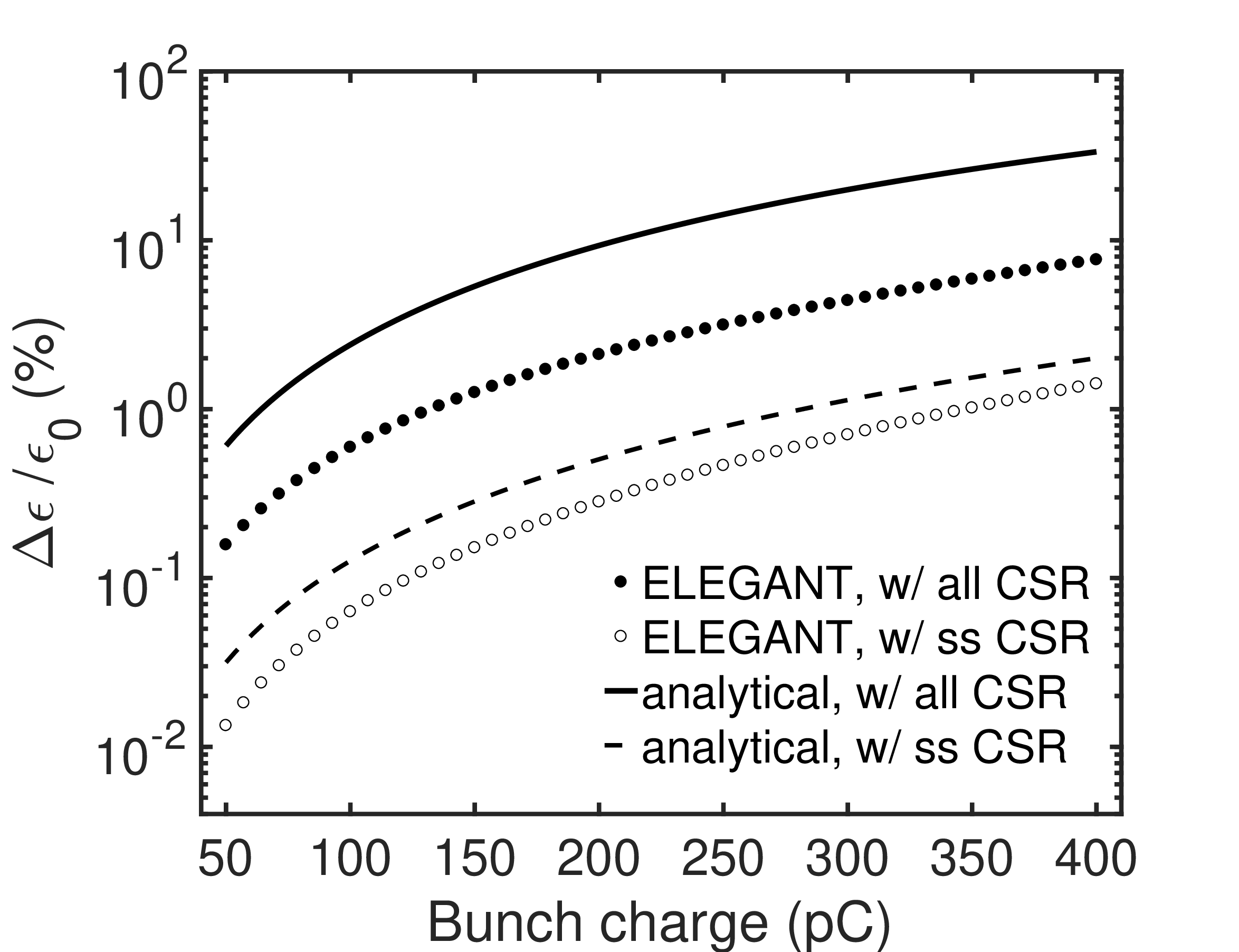}
\caption{\textmd{Relative emittance growth (in \%) versus the bunch charge for the non-symmetric S-shape chicane. The solid and dashed lines are calculated from the analytical formulas for the all CSR and steady-state CSR, respectively. The solid and empty circles are collected from particle tracking simulations for the all CSR and steady-state CSR, respectively. }}
\label{Fig6}
\end{figure}

Having looked at the MBI gain, we now study the projected emittance growth after the beam passes through the non-symmetric S-shape chicane. Figure~\ref{Fig6} shows the relative emittance growth at the exit of the lattice for different bunch charges. Two cases are considered: all CSR (steady-state and transients) and steady-state CSR only. The results calculated based on the analytical formula are given by the solid and dashed lines in the figure, respectively, while the results based on the ELEGANT simulations are represented by the solid and empty circles. Regarding the simulation parameters used for ELEGANT, here in this example we used $10^6$ simulation particles to represent a bunch. In the bending magnet, when using the element CSRCSBEND, the propagation was split into 10 parts (N\_KICKS), and the histogram bins required for each calculation of the CSR field was set to 1000 (N\_BINS). The remaining numerical setup is the same as the first example.

It can be seen from the figure that the resulting trends from the analytical formula and the tracking simulation are basically consistent, although the theoretical formula prediction is still relatively conservative, i.e., over-estimated.

\begin{figure}[htbp]
\centering
\includegraphics[width=14cm]{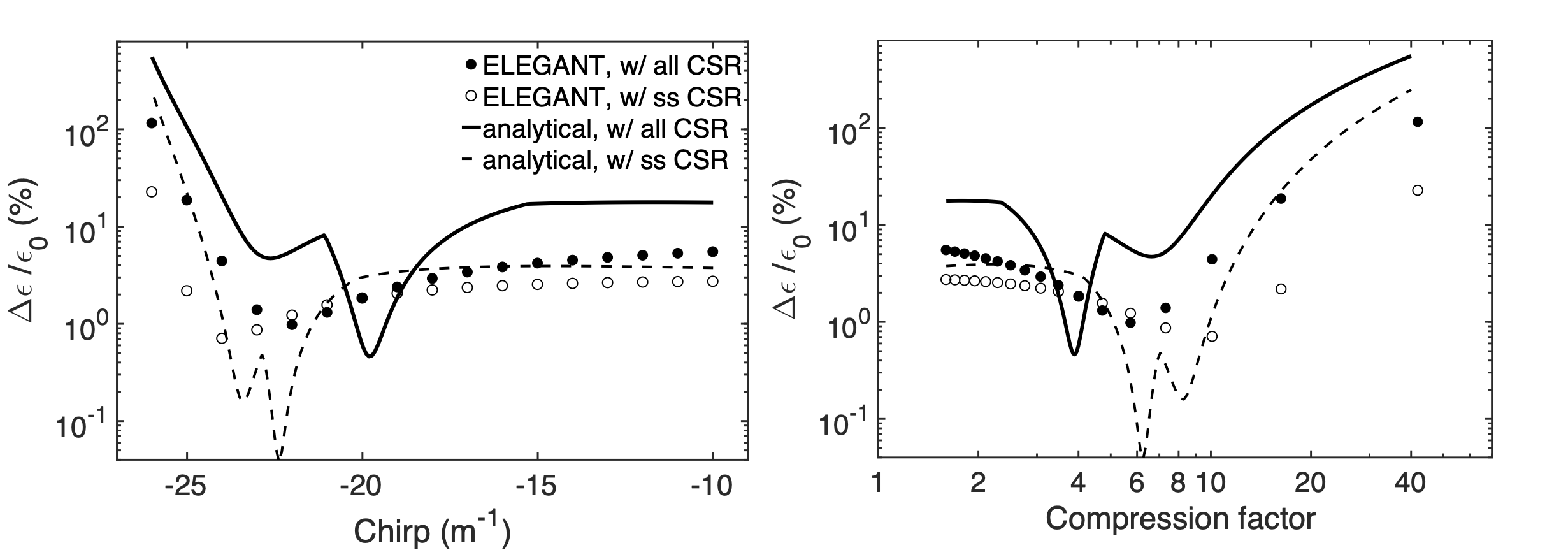}
\caption{\textmd{Relative emittance growth (in \%) as a function of the initial energy chirp and the corresponding compression factor for the non-symmetric S-chicane. The solid and dashed lines are calculated by the analytical formulas for the all CSR (including steady-state and transient CSR) and for the only steady-state CSR, respectively. The solid and empty circles represent the ELEGANT tracking results.}}
\label{Fig7}
\end{figure}

Figure~\ref{Fig7} shows the emittance growth for different initial energy chirps, which correspond to different bunch compression factors. In general, the results given by the analytical formula are slightly higher than the tracking results, which means that the analytical predictions are relatively conservative but the trend is basically consistent. Part of the reason for some discrepancy is that the $R_{16}$ and $R_{26}$ used in the analytical emittance calculation are based on the first-order approximation, and the influence of higher-order terms is ignored. The higher order effects may be reflected in CSR-induced phase space distortion. Another part of the difference is due to the value of the bunch length used when measuring the CSR-induced energy spread increase. When calculating the energy spread contributed inside each bending magnets, it is necessary to first calculate the bunch length at the entrance and exit of the section. When evaluating the bunch compression factor, due to the $R_{56}$ in the denominator of the compression factor expression, the difference in the value will also lead to the difference in the final compression factor. These indirectly affect the estimation of the energy spread, and will also cause some differences in the final evaluation of the emittance growth. The analytical results in Fig.~\ref{Fig7} are obtained using the accurate form of Eq. (\ref{Eq15}). 

Before ending this section, we would like to point out again that the second lattice example is an innovative design of a non-symmetric S-shape chicane recently proposed. We have only discussed some of the CSR-induced MBI properties and the emittance growth, as an application of the derived analytical formulas. We can see that when all CSR fields are considered, the maximum emittance growth predicted by ELEGANT is still within 10\% after the bunch compression factor of 10 with a single bunch charge of 400 pC (or an initial peak current of 5 kA)! The prediction of the analytical formula is within 30\%, on the same order. Preliminary studies~\cite{Ref36,Ref37} have shown that after sacrificing the symmetry of the C-shape chicane, such a non-symmetric chicane design seems to be able to suppress the CSR-induced emittance growth and MBI to a greater extent. For more discussions on S-shape chicane, please refer to our recently submitted paper~\cite{Ref36,Ref37}.

\section{Summary and Conclusion}\label{SecIV}

In this paper, we have derived analytical formulas for the MBI gain and the emittance growth caused by both the steady-state and transient CSR effects in an arbitrary achromatic four-bend transport line. Specifically, the gain formula is given by Eq. (\ref{Eq28}), while the analytical formulas for the emittance growth are given by Eqs. (\ref{Eq29}) and (\ref{Eq32}). We use a typical symmetric C-shaped chicane as a first example to verify the consistency between the results of the analytical formulas and particle tracking simulations. Subsequently, as the second example, we apply the formulas to the non-symmetric S-shaped chicane to estimate the CSR effects.

The main contribution of this paper is that it is the first to provide the MBI gain formula that includes both the steady-state and transient CSR effects. Formulas considering only the steady-state may not fully capture the gain amplification, as transient CSR also contributes. However, when designing a multi-bend transport lattice, our experiences from simulation results suggest that, as long as the steady-state is properly handled, the result is generally not too far off. We are also aware that the transient CSR impedance used in this paper is not complete, and its validity is limited to high energy beams.

The advantage of using analytical formulas, in addition to significantly improved computational efficiency, is the ability to study the functional dependences of the gain on various parameters. Without the need for time-consuming, full numerical particle tracking simulations, the analytical formulas can provide initial guidance for fast optimization, while considering more comprehensive CSR effects. Of course, more accurate and realistic evaluations will still rely on particle tracking numerical simulations, and our formulas could serve as the first step in the investigation.

\section{Acknowledgements}
This work is supported by the Fundamental Research Funds for the Central Universities (HUST) under Project No. 2021GCRC006 and National Natural Science Foundation of China under project No. 12275094. One of the authors (C.T.) acknowledges O.J. Tsai for help in the preparation of this manuscript.


\appendix

\section{Detailed derivation of Eq. (\ref{Eq26})}\label{AppA}

Here we provide the explicit expressions of individual integrals in Eq. (\ref{Eq26}). First, the pure optics term, independent of the current, is given by
\begin{equation}
b_0\left(k_f ; s_f\right)=\exp \left[-\frac{\bar{\sigma}_\delta^2}{2\left(1+h R_{56}\right)^2}\right] b_0\left(k_0 ; s_0=0\right),
\end{equation}
with $\bar{\sigma}_\delta=k_0 R_{56} \sigma_\delta$ and $k_f=k_0 C(s_f)=k_0 /\left(1+h R_{56}\right)$. 

For the remaining integrals, for clarity, we will organize the terms from different orders of iteration into the following separate subsections. There are three subsections, corresponding to integrals involving the first, second, and third iterations, respectively.

\subsection{Integrals involving the first iteration}

The first-order iteration terms, that are linearly dependent on the current, include three terms. The CSR-induced energy modulation in the first bending magnet and transported to the lattice exit has the following contribution
\begin{equation}
\begin{aligned}
\int_0^{L_{b 1}} d s_1 K\left(s_1, s_f\right) b_0\left[k\left(s_1\right) ; s_1\right] &= A \overline{I}_{f 1}\left( F_0+\frac{1-e^{-\bar{\sigma}_{x 1}^2}}{2 \bar{\sigma}_{x 1}^2}\right) \\
&\times \exp \left[-\frac{\bar{\sigma}_\delta^2}{2\left(1+h R_{56}\right)^2}\right] b_0\left(k_0 ; 0\right),
\end{aligned}
\end{equation}
where, as the shorthand notations, the scaled current and beam size are
\begin{equation}
\overline{I}_{f 1}=\frac{I_f k_0^{4 / 3} R_{56} L_{b 1}}{\gamma I_A \rho_1^{2 / 3}}, \quad \overline{\sigma}_{x 1}=k_0 \frac{L_{b 1} \sqrt{\epsilon_{x0} \beta_{x0}} }{ \rho_1},
\end{equation}
and
\begin{equation}
F_0\left(\bar{\sigma}_{x 1}\right)=\frac{e^{-\bar{\sigma}_{x 1}^2+\bar{\sigma}_{x 1} \sqrt{\pi} \operatorname{erf}\left(\bar{\sigma}_{x 1}\right)-1}}{2 \bar{\sigma}_{x 1}^2},
\end{equation}
here the error function is defined as $\operatorname{erf}(x)=2 \pi^{-1 / 2} \int_0^x d t \exp \left(-t^2\right)$. In the second bending magnet, the CSR-induced energy modulation may convert to density modulation through the transport to the lattice exit
\begin{equation}
\int_0^{L_{b 2}} d s_2 K\left(s_2, s_{f}\right) b_0\left[k\left(s_2\right) ; s_2\right]=A \overline{I}_{f 2} ~F_1 b_0\left(k_0 ; 0\right),
\end{equation}
with
\begin{equation}
F_1=\int_0^{d_1} d t \frac{1-t}{\left(1+h R_{56} t\right)^{\frac{4}{3}}} H_1(t),
\end{equation}
where
\begin{equation}
H_1(t)=\exp \left[-\bar{\sigma}_{x 2}^2 \frac{\left[\left(1-\alpha_{x0} \varphi_1\right) t-m_1\right]^2+\varphi_1^2 t^2}{\left(1+h R_{56} t\right)^2}-\frac{\bar{\sigma}_\delta^2}{2\left(1+h R_{56} t\right)^2}\left(t^2+\frac{(1-t)^2}{\left(1+h R_{56}\right)^2}\right)\right].
\end{equation}
In the above expressions, some shorthand notations are defined as follows
\begin{equation}
\overline{I}_{f2}=\frac{I_f k_0^{4 / 3} R_{56} \overline{L_{b 2}}}{\gamma I_A \rho_2^{2 / 3}}, \quad \bar{\sigma}_{x 2}=k_0 \frac{\overline{L_{b 2}} \sqrt{\epsilon_{x0} \beta_{x0}} }{ \rho_2}, \quad \overline{L_{b 2}}=\frac{\rho_1 \rho_2 R_{56}}{L_{b1} L_{d1}}, \quad \varphi_1=\frac{L_{d 1}}{\beta_{x0}},
\end{equation}
and $m_1=\left(\frac{L_{b 1}}{\rho_1}\right)^2 \frac{L_{d 1}}{R_{56}}, \quad d_1=\frac{L_{b 2}}{\overline{L_{b 2}}}=\frac{L_{b 1} L_{b 2} L_{d 1}}{\rho_1 \rho_2 R_{56}}$. Similarly, the CSR-induced energy modulation in the third bending magnet may convert to density modulation through the transport to the lattice exit as
\begin{equation}
\int_0^{L_{b 3}} d s_3 K\left(s_3, s_{f}\right) b_0\left[k\left(s_3\right) ; s_3\right]=A \overline{I}_{f 3} ~F_2 ~b_0\left(k_0 ; 0\right),
\end{equation}
where
\begin{equation}
F_2=\int_0^1 d t \frac{t-1}{\left\{1+h\left[\left(R_{56}-w\right) t+w\right]\right\}^{\frac{4}{3}}} H_2(t),
\end{equation}
and
\small
\begin{equation}
H_2(t)=\exp \left\{-\bar{\sigma}_{x 3}^2 \frac{\left[\left(1-\alpha_{x0} \varphi_2\right) t-m_2\right]^2+\left(m_3-\varphi_2 t\right)^2}{\left(1+h\left[\left(R_{56}-w\right) t+w\right]\right)^2}-\bar{\sigma}_\delta^2 \frac{\left[\frac{R_{56}-\left(R_{56}-w\right) t-w}{\left(1+h R_{56}\right) R_{56}}\right]^2+\left[\frac{\left(R_{56}-w\right) t+w}{R_{56}}\right]^2}{2\left(1+h\left[\left(R_{56}-w\right) t+w\right]\right)^2}\right\},
\end{equation}
\normalsize
where the shorthand notations associated with this integral are
\begin{equation}
\overline{I}_{f 3}=\frac{\left.I_f k_0^{\frac{4}{3}}\left[R_{56}\left(s_{2 f}\right)-R_{56}\right)\right] L_{b 3}}{\gamma I_A \rho_3^{2 / 3}}, \quad \overline{\sigma}_{x 3}=k_0 \frac{{L}_{b 3} \sqrt{\epsilon_{x0} \beta_{x0}}}{ \rho_3}, \quad \varphi_2=\frac{L_{d 1}+L_{d 2}}{\beta_{x0}},
\end{equation}
and
\begin{equation}
m_2=\xi\left(\frac{\rho_3 L_{b 2}}{\rho_2 L_{b 3}}-\frac{\rho_3 L_{b 1}}{\rho_1 L_{b 3}}-\frac{\alpha_{x0}}{\beta_{x0}} \frac{\rho_3 L_{b 2} L_{d 1}}{\rho_2 L_{b 3}}\right), \quad m_3=\xi \frac{\rho_3 L_{b 2} L_{d 1}}{\rho_2 L_{b 3} \beta_{x0}}, \quad w=\frac{L_{b 1} L_{b 2} L_{d 1}}{\rho_1 \rho_2}=R_{56}\left(s_{2 f}\right).
\end{equation}
In the above expressions, the parameter $\xi$ is used to differentiate the polarity of the third bending magnet, which is different for the S-shape and C-shape chicanes
\begin{equation}
\xi= \begin{cases}1, & \text { if } \frac{L_{b 1}}{\rho_1} < \frac{L_{b 2}}{\rho_2}, \quad \text{S-shape chicane} \\ -1, & \text { if } \frac{L_{b 1}}{\rho_1} \ge \frac{L_{b 2}}{\rho_2}, \quad \text{C-shape chicane} \end{cases}
\end{equation}

The contribution of the fourth bending magnet to the final gain can be neglected, because the conversion from energy to density modulation only occurs inside the bend.

\subsection{Integrals involving the second iteration}

The second-order iteration, which is proportional to the square of the bunch current, also comprises three terms. Following the energy modulation generated in the first bending magnet, the beam undergoes a free-space drift to the second magnet, where further energy modulation occurs due to CSR. Ultimately, the density modulation is converted at the lattice exit. This process is formulated as
\begin{equation}\label{EqA15}
\int_0^{L_{b 2}} d s_2 K\left(s_2, s_f\right) \int_0^{L_{b 1}} d s_1 K\left(s_1, s_2\right) b_0\left[k\left(s_1\right) ; s\right] \approx A^2 \overline{I}_{f 1} \overline{I}_{f 2} ~F_0 F_3 b_0\left(k_0 ; 0\right),
\end{equation}
where
\begin{equation}
F_3 = \int_0^{d_1} d t \frac{(1-t) t\left(1+h R_{56}\right)}{\left(1+h R_{56} t\right)^{\frac{7}{3}}} H_1(t).
\end{equation}

Here we comment that in Eq. (\ref{EqA15}) we have neglected the terms containing the product of $s_1 s_2$, e.g., the term in the exponential of the integrand
\begin{equation}
-k_0^2 \epsilon_{x0} \beta_{x0}\left(\frac{\frac{s_2}{\rho_2}-\frac{L_{b 1}}{\rho_1}}{1+h R_{56} \frac{s_2}{\overline{L_{b 2}}}}-\frac{\alpha_{x0}}{\beta_{x0}} \frac{\frac{L_{d 1} s_2}{\rho_2}}{1+h R_{56} \frac{s_2}{\overline{L_{b 2}}}}\right) \frac{s_1}{\rho_1}
\end{equation}
To reach a simplified expression for the final result, by neglecting this cross term we leave the integrations variables separated in the double integral, transforming it into a sum or difference of multiple single integrals. Upon comparison with semi-analytical Vlasov results, we find that the deviations resulting from such simplification are passable. In Ref.~\cite{Ref09}, the same simplification is applied.

In addition to the above route, there is another route to accumulate the final gain. That is, following the energy modulation induced in the first bending magnet, the beam undergoes a free-space drift to the third magnet, where further energy modulation occurs. Then the density modulation is observed at the lattice exit. This process is formulated as
\begin{equation}
\begin{aligned}
\int_0^{L_{b 3}} d s_3 K\left(s_3, s_f\right) &\int_0^{L_{b 1}} d s_1 K\left(s_1, s_3\right) b_0\left[k\left(s_1\right) ; s_1\right] \approx A^2 \overline{I}_{f 1} \overline{I}_{f 3} \times \\
&\left[F_4 \times \left(F_0+\frac{1-e^{-\bar{\sigma}_{x 1}^2}}{2 \bar{\sigma}_{x 1}^2}\right)-F_5 \frac{1-e^{-\bar{\sigma}_{x 1}^2}}{2 \bar{\sigma}_{x 1}^2}\right] b_0\left(k_0 ; 0\right),
\end{aligned}
\end{equation}
where
\begin{equation}
\begin{aligned}
F_4 & =\int_0^1 d t \frac{(t-1)\left[\left(R_{56}-w\right) t+w\right]\left(1+h R_{56}\right)}{\left\{1+h\left[\left(R_{56}-w\right) t+w\right]\right\}^{7/3} R_{56}} H_2(t), \\
F_5 & =\int_0^1 d t \frac{(t-1)\left[m_4 t+w\right]\left(1+h R_{56}\right)}{\left\{1+h\left[\left(R_{56}-w\right) t+w\right]\right\}^{7/3} R_{56}} H_2(t),
\end{aligned}
\end{equation}
with $m_4=R_{56}-w-\xi \frac{L_{b 2} L_{b 3} L_{d 2}}{\rho_2 \rho_3}$. The last term involving the second-order iteration corresponds to the scenario: the energy modulation generated in the second bending magnet, during free-space drift to the third magnet, where the beam undergoes further energy modulation, and finally resulting in density modulation at the lattice exit, contributes the following term
\begin{equation}
\int_0^{L_{b 3}} d s_3 K\left(s_3, s_f\right) \int_0^{L_{b 2}} d s_2 K\left(s_2, s_3\right) b_0\left[k\left(s_2\right) ; s_2\right] \approx A^2 \overline{I}_{f 2} \overline{I}_{f 3} ~F_6 ~F_7 ~b_0\left(k_0 ; 0\right),
\end{equation}
in which
\begin{equation}
\begin{aligned}
F_6 &= \int_0^{d_1} d t \frac{\left(1-\frac{t}{d_1}\right)}{\left(1+h R_{56} t\right)^{4/3}} H_1(t), \\
F_7 &= \int_0^1 d t \frac{t(t-1)\left(1+h R_{56}\right)\left(R_{56}-w-m_4\right)}{\left\{1+h\left[\left(R_{56}-w\right) t+w\right]\right\}^{7/3} R_{56}} H_2(t).
\end{aligned}
\end{equation}

\subsection{Integrals involving the third iteration}

Finally, the term, that is cubic in the current, corresponds to the third-order iteration, with the following physical picture: the energy modulation generated in the first bending magnet is free-space transported to the second magnet, where it undergoes further energy modulation, then transported to the third magnet for another energy modulation, and ultimately results in the density modulation at the lattice exit. This process is formulated as
\begin{equation}
\begin{aligned}
\int_0^{L_{b 3}} d s_3 K\left(s_3, s_f\right) \int_0^{L_{b 2}} d s_2 K\left(s_2, s_3\right) &\int_0^{L_{b 1}} d s_1 K\left(s_1, s_2\right) b_0\left[k\left(s_1\right) ; s_1\right] \\
&\approx A^3 \overline{I}_{f 1} \overline{I}_{f 2} \overline{I}_{f 3} ~F_0 ~F_7 ~F_8 ~b_0(k 0 ; 0),
\end{aligned}
\end{equation}
where
\begin{equation}
F_8=\int_0^{d_1} d t \frac{t\left(1-\frac{t}{d_1}\right)\left(1+h R_{56}\right)}{\left(1+h R_{56} t\right)^{7/3}} H_1(t).
\end{equation}

The terms mentioned above are the contributions to the change in the bunching factor arising from the steady-state CSR impedance.

\subsection{Integrals involving the drift CSR}

Taking into account the exit-transient CSR impedance, its contribution is reflected in the following two terms in the integrals. The energy modulation induced by the CSR field in the first drift section, transported to the lattice exit, gives
\scriptsize
\begin{equation}
\begin{aligned}
D_1 &= \int_0^{L_{d 1}} d \zeta_1 K\left(\zeta_1, s_f\right) b_0\left[k\left(\zeta_1\right) ; \zeta_1\right]  \\
&= \begin{cases}\ln \left(\frac{L_{d 1} k_0^{\frac{1}{3}}}{\rho_1^{\frac{2}{3}}(2 \pi)^{\frac{1}{3}}}\right) b_0(k_0;0) \times \\
\left(i \frac{2 k_0 I_f R_{56}}{\gamma I_A} \exp \left[-k_0^2\left(\frac{R_{56}}{1+h R_{56}}\right)^2 \frac{\sigma_\delta^2}{2}-k_0^2 \epsilon_{x0} \beta_{x0}\left(\frac{L_{b 1}}{\rho_1}-\frac{\alpha_{x0}}{\beta_{x0}} \frac{L_{b 1}^2}{2 \rho_1}\right)^2-\frac{k_0^2 \epsilon_{x0}}{\beta_{x0}} \frac{L_{b 1}^4}{4 \rho_1^2}\right]\right), & \text { if } \frac{\gamma^2}{k_0} \ge L_{d 1} \\
\left[\ln \left(\frac{\gamma^2}{\rho_1^{\frac{2}{3}} k_0^{\frac{2}{3}}(2 \pi)^{\frac{1}{3}}}\right)+\frac{k_0 L_{d 1}}{\gamma^2}-1\right] b_0(k_0;0) \times \\
\left(i \frac{2 k_0 I_f R_{56}}{\gamma I_A} \exp \left[-k_0^2\left(\frac{R_{56}}{1+h R_{56}}\right)^2 \frac{\sigma_\delta^2}{2}-k_0^2 \epsilon_{x0} \beta_{x0}\left(\frac{L_{b 1}}{\rho_1}-\frac{\alpha_{x0}}{\beta_{x0}} \frac{L_{b 1}^2}{2 \rho_1}\right)^2-\frac{k_0^2 \epsilon_{x0}}{\beta_{x0}} \frac{L_{b 1}^4}{4 \rho_1^2}\right]\right), & \text { if } \frac{\gamma^2}{k_0}<L_{d 1}\end{cases}
\end{aligned}
\end{equation}
\normalsize

The CSR-induced energy modulation induced in the second drift section, transported to the lattice exit, contributes the following term
\tiny
\begin{equation}
\begin{aligned}
D_2 &= \int_0^{L_{d2}} d \zeta_2 K\left(\zeta_2, s_f\right) b_0\left[k\left(\zeta_2\right) ; \zeta_2\right]  \\
&=\left\{\begin{array}{l}
\ln \left(\frac{L_{d 2} k_0^{\frac{1}{3}}}{\rho_2^{\frac{2}{3}}(2 \pi)^{\frac{1}{3}}(1+h w)^{\frac{1}{3}}}\right) \times i \frac{2 k_0 I_f\left(R_{56}-w\right)}{\gamma I_A(1+h w)} \times b_0(k_0;0) \times \\
\exp \left[-k_0^2\left(\frac{R_{56}}{1+h R_{56}}-\frac{w}{1+h w}\right)^2 \frac{\sigma_\delta^2}{2}-k_0^2 \epsilon_{x0} \beta_{x0}\left(\frac{1}{1+h w}\left(\frac{L_{b 1}}{\rho_1}-\frac{L_{b 2}}{\rho_2}\right)+\frac{\alpha_{x0}}{\beta_{x0}} \frac{L_{b 2} L_{d 1}}{(1+h w) \rho_2}\right)^2-\frac{k_0^2 \epsilon_{x0}}{\beta_{x0}}\left(\frac{L_{b 2} L_{d 1}}{(1+h w) \rho_2}\right)^2\right], ~\text { if } \frac{(1+hw)\gamma^2}{k_0} \ge L_{d 2} \\
\left[\ln \left(\frac{(1+h w)^{\frac{2}{3}} \gamma^2}{\rho_2^{\frac{2}{3}} k_0{ }^{\frac{2}{3}}(2 \pi)^{\frac{1}{3}}}\right)+\frac{k_0 L_{d 2}}{\gamma^2(1+h w)}-1\right] \times i \frac{2 k_0 I_f\left(R_{56}-w\right)}{\gamma I_A(1+h w)} \times b_0(k_0;0) \times \\
\exp \left[-k_0^2\left(\frac{R_{56}}{1+h R_{56}}-\frac{w}{1+h w}\right)^2 \frac{\sigma_\delta^2}{2}-k_0^2 \epsilon_{x0} \beta_{x0}\left(\frac{1}{1+h w}\left(\frac{L_{b 1}}{\rho_1}-\frac{L_{b 2}}{\rho_2}\right)+\frac{\alpha_{x0}}{\beta_{x0}} \frac{L_{b 2} L_{d 1}}{(1+h w) \rho_2}\right)^2-\frac{k_0^2 \epsilon_{x0}}{\beta_{x0}}\left(\frac{L_{b 2} L_{d 1}}{(1+h w) \rho_2}\right)^2\right], ~\text { if } \frac{(1+hw)\gamma^2}{k_0} < L_{d 2}
\end{array}\right. \\
\end{aligned}
\end{equation}
\normalsize
with $w=\frac{L_{b 1} L_{b 2} L_{d 1}}{\rho_1 \rho_2}=R_{56}\left(s_{2 f}\right)$.

In the third drift section, since $R_{56}(\zeta_3 \to s_f) \approx 0$, we have
\begin{equation}
D_3 = \int_0^{L_{d 3}} d \zeta_3 K\left(\zeta_3, s_f\right) b_0\left[k\left(\zeta_3\right) ; \zeta_3\right] \approx 0,
\end{equation}
that is, the energy modulation induced by the exit-transient CSR field does not contribute significantly to the final density modulation gain and can be neglected. Summing up the above results, we obtain the total gain formula in Eq. (\ref{Eq28}).

Here we comment that for the case including both the steady-state and the exit-transient CSR, when substituting the exit-transient CSR impedance into the integral equation, we only consider the first-order iteration to simplify the final form of the MBI gain formula. Furthermore, in computing with the analytical formula, we find that the final gain magnitude may depend sensitively on the value of the overall $R_{56}$ of the transport line. Mathematically, this is because $R_{56}$ appears frequently in the numerator or denominator of the integrals in the analytical formula, especially the denominator, and a slight difference in the value of $R_{56}$ can potentially affect the final gain. Physically, a slightly different value of $R_{56}$ implies a different bunch compression process, and sensitively during large compression factor, involving real-time changes in the beam current during the transport, leading to different degrees of microbunching instability and thus affecting the final gain result.

\section{Detailed derivation of Eqs. (\ref{Eq32})}\label{AppB}

Here we provide the explicit forms of each term in Eqs. (\ref{Eq32}). Based on the optical functions summarized in the main text (Sec.~\ref{SecIIA}), and after lengthy but straightforward integrations, we obtain the analytical formulas for the emittance growth in the S-shaped and C-shaped chicanes as follows. Following the same convention introduced earlier
\begin{equation}
\xi=\left\{\begin{array}{lll}
1, & \text { if } \frac{L_{b 1}}{\rho_1}<\frac{L_{b 2}}{\rho_2}, & \text { S-shape chicane } \\
-1, & \text { if } \frac{L_{b 1}}{\rho_1} \geq \frac{L_{b 2}}{\rho_2}, & \text { C-shape chicane }
\end{array}\right.
\end{equation}

To evaluate the emittance growth due to CSR, we first calculate the CSR-induced energy spread increase using the formulas from Sec.~\ref{SecIIB}, assuming a 1-D Gaussian line bunch. Then, we integrate the coherent transverse beam size and angular divergence through the following expressions
\begin{equation}
\begin{aligned}
\left\langle\Delta x^2\right\rangle_{\mathrm{coh}} & =\left(\int_{s_0}^{s_f} R_{16}\left(s \rightarrow s_f\right) \frac{\partial \sigma_{\delta, \mathrm{CSR}}}{\partial s} d s\right)^2, \\
\left\langle\Delta x^{\prime 2}\right\rangle_{\mathrm{coh}} & =\left(\int_{s_0}^{s_f} R_{26}\left(s \rightarrow s_f\right) \frac{\partial \sigma_{\delta, \mathrm{CSR}}}{\partial s} d s\right)^2,
\end{aligned}
\end{equation}
where in the integrand the change of energy spread per unit length is approximated as
\begin{equation}
\frac{\partial \sigma_{\delta, \mathrm{CSR}}}{\partial s} \approx \frac{\Delta \sigma_{\delta, \mathrm{CSR}}}{L_b \text { or } L_d}.
\end{equation}

By substituting the results from Eqs. (\ref{Eq18}) and (\ref{Eq20}) into Eqs. (\ref{Eq32}), and obtaining $\left\langle\Delta x^2\right\rangle_{\mathrm{coh}}$ and $\left\langle\Delta x^{\prime 2}\right\rangle_{\mathrm{coh}}$, the respective terms in the emittance increase formula are given as follows
\begin{align}
g_1 &= \int_{0}^{L_{b1}} R_{16}\left(s_1 \rightarrow s_f\right) \frac{\partial \sigma_{\delta, \mathrm{CSR}}}{\partial s} d s \nonumber \\
&= - \int_0^{L_{b 1}}\left[R_{16}\left(s_1\right)+\left(L_T-s_1\right) R_{26}\left(s_1\right)\right] \frac{\partial \sigma_{\delta B 1}}{\partial s_1} d s_1 \nonumber \\
&\approx - \frac{\Delta \sigma_{\delta B 1}}{L_{b1}} \int_0^{L_{b 1}}\left[R_{16}\left(s_1\right)+\left(L_T-s_1\right) R_{26}\left(s_1\right)\right] d s_1 = - \frac{3 L_T L_{b 1}-L_{b 1}^2}{6 \rho_1} \Delta \sigma_{\delta B 1}\\
g_2 &= -\int_0^{L_{d 1}}\left[R_{16}\left(\zeta_1\right)+\left(L_1-\zeta_1\right) R_{26}\left(\zeta_1\right)\right] \frac{\partial \sigma_{\delta D 1}}{\partial \zeta_1} d \zeta_1 \approx - \frac{L_1 L_{b 1}}{\rho_1} \Delta \sigma_{\delta D 1} \\
g_3 &= -\int_0^{L_{b 2}}\left[R_{16}\left(s_2\right)+\left(L_2-s_2\right) R_{26}\left(s_2\right)\right] \frac{\partial \sigma_{\delta B 2}}{\partial s_2} d s_2 \nonumber \\
&\approx - \left(\frac{L_{b 1} L_1}{\rho_1}-\frac{L_{b 2} L_2}{2 \rho_2}-\frac{L_{b 1} L_{b 2}}{2 \rho_1}+\frac{L_{b 2}^2}{3 \rho_2}\right) \Delta \sigma_{\delta B 2} \\
g_4 &= -\int_0^{L_{d 2}}\left[R_{16}\left(\zeta_2\right)+\left(L_3-\zeta_2\right) R_{26}\left(\zeta_2\right)\right] \frac{\partial \sigma_{\delta D 2}}{\partial \zeta_2} d \zeta_2 \nonumber \\
& \approx - \left(\frac{L_{b 1} L_{d 1}}{\rho_1}+\frac{L_{b 1} L_3}{\rho_1}-\frac{L_{b 2} L_3}{\rho_2}\right) \Delta \sigma_{\delta D 2}
\end{align}
with $L_1=L_T-L_{b 1}$, $L_2=L_1-L_{d 1}$, and $L_3=L_2-L_{b 2}$. Moreover,
\begin{align}
g_5 &= -\int_0^{L_{b 3}}\left[R_{16}\left(s_3\right)+\left(L_4-s_3\right) R_{26}\left(s_3\right)\right] \frac{\partial \sigma_{\delta B 3}}{\partial s_3} d s_3 \nonumber \\
&\approx - \left[\left(\frac{L_{b 1} L_{d 1}}{\rho_1}+\frac{L_{b 1} L_{d 2}}{\rho_1}-\frac{L_{b 2} L_{d 2}}{\rho_2}\right)+\left(L_4-\frac{L_{b 3}}{2}\right)\left(\frac{L_{b 1}}{\rho_1}-\frac{L_{b 2}}{\rho_2}\right)+\xi \frac{L_4 L_{b 3}}{2 \rho_3}-\xi \frac{L_{b 3}^2}{3 \rho_3}\right] \Delta \sigma_{\delta B 3} \\
g_6 &= -\int_0^{L_{d 3}}\left[R_{16}\left(\zeta_3\right)+\left(L_5-\zeta_3\right) R_{26}\left(\zeta_3\right)\right] \frac{\partial \sigma_{\delta D 3}}{\partial \zeta_3} d \zeta_3 \nonumber \\
&\approx - \left[\left(\frac{L_{b 1} L_{d 1}}{\rho_1}+\frac{L_{b 1} L_{d 2}}{\rho_1}-\frac{L_{b 2} L_{d 2}}{\rho_2}\right)+L_5\left(\frac{L_{b 1}}{\rho_1}-\frac{L_{b 2}}{\rho_2}+\xi \frac{L_{b 3}}{\rho_3}\right)\right] \Delta \sigma_{\delta D 3} \\
g_7 &= -\int_0^{L_{b 4}}\left[R_{16}\left(s_4\right)+\left(L_{b 4}-s_4\right) R_{26}\left(s_4\right)\right] \frac{\partial \sigma_{\delta B 4}}{\partial s_4} d s_4 \approx - \xi \frac{L_{b 4}^2}{3 \rho_4} \Delta \sigma_{\delta B 4}
\end{align}
with $L_4=L_3-L_{d 2}$ and $L_5=L_4-L_{b 3}$. Those terms associated with the change of transverse beam divergence include
\begin{align}
g_8 & =-\int_0^{L_{b 1}} R_{26}\left(s_1\right) \frac{\partial \sigma_{\delta B 1}}{\partial s_1} d s_1 \approx - \frac{\Delta \sigma_{\delta B 1}}{L_{b1}} \int_0^{L_{b 1}} R_{26}\left(s_1\right) d s_1 = - \frac{L_{b 1}}{2 \rho_1} \Delta \sigma_{\delta B 1} \\
g_9 &= -\int_0^{L_{d 1}} R_{26}\left(\zeta_1\right) \frac{\partial \sigma_{\delta D 1}}{\partial \zeta_1} d \zeta_1 \approx - \frac{L_{b 1}}{\rho_1} \Delta \sigma_{\delta D 1} \\
g_{10} &= -\int_0^{L_{b 2}} R_{26}\left(s_2\right) \frac{\partial \sigma_{\delta B 2}}{\partial s_2} d s_2 \approx - \left(\frac{L_{b 1}}{\rho_1}-\frac{L_{b 2}}{2 \rho_2}\right) \Delta \sigma_{\delta B 2} \\
g_{11} &= -\int_0^{L_{d 2}} R_{26}\left(\zeta_2\right) \frac{\partial \sigma_{\delta D 2}}{\partial \zeta_2} d \zeta_2 \approx - \left(\frac{L_{b 1}}{\rho_1}-\frac{L_{b 2}}{\rho_2}\right) \Delta \sigma_{\delta D 2} \\
g_{12} &= -\int_0^{L_{b 3}} R_{26}\left(s_3\right) \frac{\partial \sigma_{\delta B 3}}{\partial s_3} d s_3 \approx - \left(\frac{L_{b 1}}{\rho_1}-\frac{L_{b 2}}{\rho_2}+\xi \frac{L_{b 3}}{2 \rho_3}\right) \Delta \sigma_{\delta B 3} \\
g_{13} &= -\int_0^{L_{d 3}} R_{26}\left(\zeta_3\right) \frac{\partial \sigma_{\delta D 3}}{\partial \zeta_3} d \zeta_3 \approx - \left(\frac{L_{b 1}}{\rho_1}-\frac{L_{b 2}}{\rho_2}+\xi \frac{L_{b 3}}{\rho_3}\right) \Delta \sigma_{\delta D 3} \\
g_{14} &= -\int_0^{L_{b 4}} R_{26}\left(s_4\right) \frac{\partial \sigma_{\delta B 4}}{\partial s_4} d s_4 \approx - \xi \frac{L_{b 4}}{2 \rho_4} \Delta \sigma_{\delta B 4}
\end{align}

For the S-shaped and C-shaped chicanes, due to the different polarities of the third and fourth bending magnets, there is a slight difference in the values of $\xi$. Here we note that, for simplicity, in deriving the increase in emittance due to CSR, we have used the first-order approximations for $R_{16}$ and $R_{26}$ and neglected the effects from higher-order terms.

\clearpage



\end{document}